# Mechanics of Cosserat media:  II.  relativistic theory

D. H. Delphenich
Spring Valley, OH 45370 USA

**Abstract.**  The derivation of the non-relativistic Cosserat equations that was described in Part I of this series of papers is extended from the group of rigid motions in three-dimensional Euclidian space to the Poincaré group of four-dimensional Minkowski space.  Examples of relativistic Cosserat media are then given in the form of the free Dirac electron and the Weyssenhoff fluid.

## Table of contents

**1. Introduction.** This article constitutes a continuation of a previous article that was concerned with a formulation of the mechanics for Cosserat media in the non-relativistic approximation. Thus, we shall assume that the reader has some familiarity with the basic line of reasoning that was detailed there. The main alterations to the previous formalism will simply involve a change in the dimension and metric signature of the space-time manifold and a change in the associated Lie group from the group of rigid motions of Euclidian 3-space to the Poincaré group of Minkowski space. Therefore, in this article we shall essentially follow a parallel course in which we replace three-dimensional Euclidian space $E^3$ with four-dimensional Minkowski space $\mathfrak{M}^4$, the group of rigid motions $ISO(3)$ with the Poincaré group $ISO(1, 3)$, and the bundle $SO(E^3) \to E^3$ of oriented, orthonormal frames on Euclidian space with the bundle $L_0(\mathfrak{M}^4) \to \mathfrak{M}^4$ of oriented, Lorentzian frames on Minkowski space. That way, we can focus more upon the aspects of the situation that are specific to the choice of Lie group, while pointing out the things that are generic to Lie groups that take the form of semi-direct products of translation groups and subgroups of the general linear group; i.e., subgroups of the affine group.

Ultimately, we shall show that some important physical examples of relativistic Cosserat media include the Dirac electron and Weyssenhoff fluids, which are relativistic, spinning fluids that serve as a simplification of the tensorial form of the Dirac equation.

**2. The Poincaré group.** Let $\mathfrak{M}^4 = (A^4, \eta)$ be four-dimensional Minkowski space, which we will regard as a four-dimensional affine space $A^4$ with a Minkowski scalar product $\eta$ on its tangent spaces; i.e., a *Lorentzian* structure.

*a. Minkowski space.* Since we are dealing with an affine space, we can identify the points of $A^4$ with the vectors in any tangent space $T_x A^4$ by associating any point $y \in A^4$ with the translation $a^\mu \in \mathbb{R}^4$ that takes $x$ to $y$ and then associating $a^\mu$ with the tangent vector $\mathbf{a} = a^\mu \, \mathbf{e}_\mu$, where $\mathbf{e}_\mu$ is a chosen linear frame in $T_x A^4$. One can then denote the translation of $x$ to $y$ as $y = x + a^\mu$, or regard it as a binary map $A^4 \times A^4 \to \mathbb{R}^4$, $(x, y) \mapsto a^\mu = y - x$. This implicitly assumes that one generally uses the canonical frame on $\mathbb{R}^4$, which is defined by:

$$\boldsymbol{\delta}_0 = (1, 0, 0, 0), \boldsymbol{\delta}_1 = (0, 1, 0, 0), \boldsymbol{\delta}_2 = (0, 0, 1, 0), \boldsymbol{\delta}_3 = (0, 0, 0, 1). \tag{2.1}$$

If one regards the frame members $\mathbf{e}_\mu$ as partial derivative operators – namely, $e_\mu^\nu \, \partial / \partial x^\nu$ – then $\mathbf{a}f$ will represent the directional derivative of any differentiable function $f$ on $A^4$ in the direction of the translation $a^\mu$.

An *affine frame* in $A^4$ amounts to an ordered pair $(O, \mathbf{e}_\mu)$, where $O$ is a point of $A^4$ that will serve as an "origin," and $\mathbf{e}_\mu$ is a linear frame in $T_O A^4$. When one is given such an affine frame, one can define a diffeomorphism of $A^4$ with $\mathbb{R}^4$ by taking any $x \in A^4$ to $x^\mu$, where $x - O = x^\mu \mathbf{e}_\mu$.

Thus, there will be some global coordinate system $\{x^\mu, \mu = 0, 1, 2, 3\}$ on $A^4$ such that the $\mathbf{e}_\mu$ take the form of the natural frame field for this coordinate system:

$$\mathbf{e}_\mu (x) = \frac{\partial}{\partial x^\mu}.$$

This frame field $\mathbf{e}_\mu (x)$ can be obtained from any affine frame $(O, \mathbf{e}_\mu)$ by translating the frame $\mathbf{e}_\mu$ at $O$ to every other tangent space $T_x A^4$ by using the differential of the left-translation map $\mathrm{L}_\mathbf{a} : A^4 \to A^4$, $O \mapsto O + a^\mu$, where $a^\mu = x - O$. Hence, the global frame field will be translation-invariant, by construction.

If $(O', \mathbf{e}'_\mu)$ is any other affine frame then since:

$$O' = O + a^\mu \mathbf{e}_\mu, \qquad\qquad \mathbf{e}'_\mu = \mathbf{e}_\nu L^\nu_\mu \qquad\qquad (2.2)$$

for some invertible 4×4 real matrix $L^\nu_\mu$ and some (unique) displacement $a^\mu$, one will see that one can express the transition from $(O, \mathbf{e}_\mu)$ to $(O', \mathbf{e}'_\mu)$ by way of the (right) action of an element $(a^\mu, L^\nu_\mu)$ of the four-dimensional affine group $A(4)$, which is the semi-direct product $\mathbb{R}^4 \times_s GL(4)$:

$$(O', \mathbf{e}'_\mu) = (O, \mathbf{e}_\mu)(a^\mu, L^\nu_\mu). \qquad\qquad (2.3)$$

The semi-direct product on $A(4)$ then takes the form:

$$(a^\mu, L^\nu_\mu)(b^\mu, M^\nu_\mu) = (a^\mu + L^\mu_\nu b^\nu, L^\mu_\kappa M^\kappa_\nu). \qquad\qquad (2.4)$$

The identity element $e$ takes the form $(0, \delta^\mu_\nu)$ and the inverse of $(a^\mu, L^\mu_\nu)$ is $(-\tilde{L}^\mu_\nu a^\nu, \tilde{L}^\mu_\mu)$, where the tilde denotes the inverse matrix, as it did in Part I.

The product on $A(4)$ can be conveniently represented as a matrix product by treating the coordinates of $\mathfrak{M}^4$ as the inhomogeneous coordinates of $\mathbb{R}\mathrm{P}^4$ and embedding $\mathfrak{M}^4$ in the space $\mathbb{R}^5 - \{0\}$ of inhomogeneous coordinates as the affine hyperplane $(1, x^\mu)$ whose first homogeneous coordinate is 1. The matrices that represent the identity matrix, $(a^\mu, L^\nu_\mu)$, and its inverse $(-\tilde{L}^\mu_\nu a^\nu, \tilde{L}^\nu_\mu)$ are then:

$$\begin{bmatrix} 1 & 0 \\ \hline 0 & \delta^\mu_\nu \end{bmatrix}, \qquad \begin{bmatrix} 1 & 0 \\ \hline a^\mu & L^\mu_\nu \end{bmatrix}, \qquad \begin{bmatrix} 1 & 0 \\ \hline -\tilde{L}^\mu_\nu a^\nu & \tilde{L}^\mu_\nu \end{bmatrix}$$

respectively.

The Minkowski space scalar product can be defined by specifying a global frame field $\{\mathbf{e}_\mu, \mu = 0, 1, 2, 3\}$ that is *Lorentzian*; i.e., it is orthonormal for $\eta$:

$$\eta(\mathbf{e}_\mu, \mathbf{e}_\nu) = \eta_{\mu\nu}, \qquad\qquad \eta = \eta^{\mu\nu}\, \mathbf{e}_\mu\, \mathbf{e}_\nu, \qquad\qquad (2.5)$$

with:

$$\eta_{\mu\nu} = \eta^{\mu\nu} = \text{diag}[+1, -1, -1, -1]. \qquad\qquad (2.6)$$

Here, we are omitting the explicit mention of the symmetrized tensor product and simply denoting it by $\mathbf{e}_\mu\, \mathbf{e}_\nu = \mathbf{e}_\nu\, \mathbf{e}_\mu$.

An essential property of the scalar product $\eta$ is that $\eta_x(\mathbf{v}, \mathbf{v})$ vanishes for a set of non-zero tangent vectors in $T_x\mathfrak{M}^4$, which one calls the *light cone at x*; such vectors are then called *light-like*. The remaining vectors then fall into two disjoint subsets: the *time-like* vectors, for which $\eta_x(\mathbf{v}, \mathbf{v}) > 0$, and the *space-like* vectors, for which $\eta_x(\mathbf{v}, \mathbf{v}) < 0$. A vector is called *causal* iff it is not space-like, and a differentiable curve in $\mathfrak{M}^4$ is called time-like (light-like, space-like, causal, resp.) iff its velocity vectors are all time-like (light-like, space-like, causal, resp.).

*a. Finite Poincaré transformations.* The relativistic analogue of rigid motions is the subgroup of the group Isom($\mathfrak{M}^4$) of all diffeomorphisms $f: \mathfrak{M}^4 \to \mathfrak{M}^4$ that preserve the scalar product, in the sense that for any $x \in \mathfrak{M}^4$ one will have:

$$\eta_{f(x)}(df|_x(\mathbf{v}), df|_x(\mathbf{w})) = \eta_x(\mathbf{v}, \mathbf{w}) \qquad\qquad (2.7)$$

for every $\mathbf{v}, \mathbf{w} \in T_x(\mathfrak{M}^4)$. This statement can be abbreviated by saying that $f^*\eta = \eta$; i.e., the pull-back of $\eta$ by $f$ is always $\eta$ again.

Relative to a coordinate system $x^\mu$ on $\mathfrak{M}^4$, this condition takes the component form:

$$\eta_{\kappa\lambda}\frac{\partial f^\kappa}{\partial x^\mu}(x)\frac{\partial f^\lambda}{\partial x^\nu}(x) = \eta_{\mu\nu}. \qquad\qquad (2.8)$$

Since one is differentiating $f$, one immediately sees that constant translations will be rigid motions, in this sense, since $f^\mu = x^\mu + a^\mu$ will give:

$$\frac{\partial f^\mu}{\partial x^\nu} = \delta^\mu_\nu, \qquad\qquad (2.9)$$

when the $a^\mu$ are constant.

Similarly, if one identifies $A^4$ with $\mathbb{R}^4$ by means of any translation-invariant global Lorentzian frame field $\mathbf{e}_\mu$ then one will see that since the differential $dL|_x$ of any linear map $L: \mathbb{R}^4 \to \mathbb{R}^4$ at any point $x$ is equivalent to a linear map from $\mathbb{R}^4$ to itself, the group Isom($\mathfrak{M}^4$) will also include the diffeomorphisms of $A^4$ that correspond to invertible linear maps $L$ of $\mathbb{R}^4$ to itself such that:

$$L^{\mathrm{T}} \eta L = \eta, \tag{2.10}$$

in which we are now referring to the matrices of everything with respect to $\mathbf{e}_\mu$. The subgroup of $GL(4)$ that is defined by all such invertible 4×4 real matrices with the latter property will be denoted by $O(1, 3)$ and referred to as the *Lorentz-orthogonal group* for $\mathfrak{M}^4$.

One can also characterize the defining property of Lorentz-orthogonal matrices by saying that:

$$L^{-1} = L^* \equiv \eta L^{\mathrm{T}} \eta, \tag{2.11}$$

and one can refer to the matrix $L^*$ as the *Lorentz-adjoint* of $L$.

One finds that $\mathrm{Isom}(\mathfrak{M}^4)$ is isomorphic to the semi-direct product $\mathbb{R}^4 \times_s O(1, 3)$, although, once again, the isomorphism is not unique and essentially comes down to a choice of inhomogeneous Lorentzian frame $(O, \mathbf{e}_\mu)$ on $\mathfrak{M}^4$, where $O$ is a point of $A^4$ that will serve as "origin" and $\mathbf{e}_\mu$ is a Lorentzian frame in $T_O A^4$. Thus, one can represent $\mathrm{Isom}(\mathfrak{M}^4)$ as a subgroup of $A(4)$.

Just as not all isometries of Euclidian 3-space represent physical motions, because one has to exclude reflections, similarly not all Lorentz-orthogonal maps represent physical motions, due to the existence of not only reflections, but time reversals.

One can first restrict to the Lorentz-orthogonal maps that also preserve a choice of volume element on $A^4$, which is then a non-zero 4-form:

$$V = dx^0 \wedge dx^1 \wedge dx^2 \wedge dx^3 = \frac{1}{4!} \, \varepsilon_{\kappa\lambda\mu\nu} \, dx^\kappa \wedge dx^\lambda \wedge dx^\mu \wedge dx^\nu. \tag{2.12}$$

Thus, one can say that the isometry $f$ preserves the volume element iff:

$$f^* V = V,$$

which translates into the component condition:

$$\det \frac{\partial f^\mu}{\partial x^\nu} = 1,$$

since, in fact:

$$f^* V_{f(x)} = \det(df|_x) \, V_x$$

for every $x$.

Once again, the translations will be included among these translations, and the linear transformations define the group $SO(1, 3)$, which we call the *proper Lorentz group*, while the semi-direct product $ISO(1, 3) \cong \mathbb{R}^4 \times_s SO(1, 3)$ will be called the *Poincaré group*.

Furthermore, due to the fact that one frame member – say, $\mathbf{e}_0$ – of a Lorentzian frame is always distinguished by being the only time-like element, while the other three $\mathbf{e}_i$, $i = 1$, 2, 3 span a space-like subspace $\Sigma_x$ of the tangent space at any point $x$, one will always have a "time+space" decomposition of $T_x \mathfrak{M}^4$ into a direct sum $[\mathbf{e}_0] \oplus \Sigma_x$, where the one-dimensional subspace $[\mathbf{e}_0]$ will be the line generated by all scalar multiples of $\mathbf{e}_0$ . One can then partition 4×4 real matrices into time-space form:

$$L_\nu^\mu = \left[\begin{array}{c|c} L_0^0 & L_0^i \\ \hline L_j^0 & L_j^i \end{array}\right] \qquad (i, j = 1, 2, 3).$$

In the case of $O(1, 3)$, the 3×3 submatrices belong to the group $O(3)$ of Euclidian rotation matrices of $E^3$, while in the case of $SO(1, 3)$, they belong to the group $SO(3)$ of (spatial) orientation-preserving Euclidian rotation matrices. For the latter group, one must have that det $L = 1$.

Since a line does not have a natural preferred orientation, but has two possible orientations $\pm [\mathbf{e}_0]$, which are defined by the directions of the positive and negative scalar multiples of $\mathbf{e}_0$, one can further define a *time orientation* on $\mathfrak{M}^4$ by choosing one or the other. Thus, one can also further restrict Lorentz-orthogonal transformations by demanding that they also preserve the time orientation:

$$f_* \pm [\mathbf{e}_0] = \pm [\mathbf{e}_0].$$

The component form of this is then:

$$\frac{\partial f^0}{\partial x^0} \delta_0^\mu = \rho \; \delta_0^\mu, \qquad \rho > 0,$$

since the components of $\mathbf{e}_0$ with respect to $\mathbf{e}_\mu$ are $(1, 0, 0, 0) = \delta_0^\mu$. More concisely, the condition is:

$$\frac{\partial f^0}{\partial x^0} > 0.$$

Translations will always have this property, but Lorentz transformations (for which det $L_\nu^\mu = 1$) will only have that property when $L_0^0 > 0$. In fact, one finds that one can decompose $O(1, 3)$ into four connected components according to the signs of det $L_\nu^\mu$ and $L_0^0$, and the component $SO_+(1, 3)$ for which both are positive is the one that contains the identity element, and is referred to as the *proper isochronous Lorentz group*. The subgroup $ISO_+(1, 3) = \mathbb{R}^4 \times_s SO_+(1, 3)$ of the Poincaré group that it defines will then be *isochronous Poincaré group*, for us.

The group $ISO_+(1, 3)$ is then a connected, ten-dimensional Lie group, although not a compact or semi-simple one. The multiplication of elements then works in a manner that is analogous to what we defined in (2.4):

$$(a^\mu, L_\nu^\mu)(\overline{a}^\mu, \overline{L}_\nu^\mu) = (a^\mu + L_\nu^\mu \, \overline{a}^\nu, L_\kappa^\mu \, \overline{L}_\nu^\kappa). \tag{2.13}$$

Since $ISO_+(1, 3)$ is a subgroup of $A^4$, it inherits the same identity element and inverse structure, so the identity element $e$ will still take the form $(0, \; \delta_\nu^\mu)$ and the inverse of $(a^\mu, L_\nu^\mu)$ will be $(-\tilde{L}_\nu^\mu a^\nu, \tilde{L}_\nu^\mu)$. One can also represent the element $(a^\mu, L_\nu^\mu)$ as an invertible 5×5 real matrix, as in the previous subsection.

The group of rigid motions of $E^3$ is contained in $ISO_+(1, 3)$ as the six-dimensional Lie subgroup of elements of the form $(a^i, \; L_j^i)$, $i, j = 1, 2, 3$, in which $L_j^i \in SO(3)$. Thus, the remaining four dimensions of $ISO_+(1, 3)$ are accounted for by the time translations, which then have the form $(\Delta t, 0, 0, 0)$, and the three elementary *boosts*, which transform any Lorentz frame to other Lorentz frame that moves with a constant relative spatial velocity with respect to the initial one. The actual decomposition of the group manifold $SO_+(1, 3)$ into a product manifold $SO(3) \times B(3)$, where $B(3)$ is the manifold of all boosts (which is not, however, a Lie group, since the multiplication of boosts does not close) is easier to discuss in the context of infinitesimal transformations, so we shall defer that until the next section.

*c. Infinitesimal Poincaré transformations.* When one differentiates a differentiable curve $(a^\mu(s), L_\nu^\mu(s))$ in $ISO_+(1, 3)$ through the identity element $e$, one will obtain an element $(v^\mu, \; \omega_\nu^\mu)$ of $T_e ISO_+(1, 3)$:

$$v^\mu = \frac{d}{ds}\bigg|_{s=0} a^\mu(s), \qquad \omega_\nu^\mu = \frac{d}{ds}\bigg|_{s=0} L_\nu^\mu(s) \qquad (2.14)$$

that can be identified with the Lie algebra $\mathfrak{iso}(1, 3)$ of $ISO_+(1, 3)$, which is defined by the left-invariant vector fields on that Lie group. Thus, $(v, \; \omega)$ represents the infinitesimal generator of a one-parameter subgroup of isochronous Poincaré transformations.

The semi-direct product in the Lie group $ISO_+(1, 3)$ corresponds to a semi-direct sum $\mathbb{R}^4 \oplus_s \mathfrak{so}(1,3)$ in $\mathfrak{iso}(1, 3)$. If $(v, \; \omega)$ and $(v', \; \omega')$ are elements of $\mathfrak{iso}(1, 3)$ then their Lie bracket will take the form:

$$[(v, \; \omega), (v', \; \omega')] = \omega v' - \omega' v + [\omega, \; \omega']. \qquad (2.15)$$

Thus, we can identify the special cases:

$$[v, \; v'] = 0, \qquad [v, \; \omega'] = -\omega' v, \qquad [\omega, \; \omega'] = \omega'. \qquad (2.16)$$

One then sees that the infinitesimal translations represent an Abelian sub-algebra, and the infinitesimal Lorentz transformations represent a non-Abelian one, while the middle expression above is what prevents the semi-direct sum from becoming a direct sum, for which the right-hand side would have to vanish.

The Lie sub-algebra $\mathfrak{so}(1, 3)$ of infinitesimal Lorentz transformations can be decomposed (as a vector space) into the direct sum $\mathfrak{so}(3) \oplus \mathfrak{b}(3)$, where $\mathfrak{so}(3)$ constitutes the Lie algebra of infinitesimal Euclidian rotations of the spatial subspace of $\mathfrak{M}^4$ and $\mathfrak{b}(3)$ is the three-dimensional vector space of infinitesimal boosts. The decomposition is the relativistic equivalent of the polarization of matrices when one replaces the matrix transpose with the Lorentz adjoint operation. One then gets:

$$\Omega = \omega + \beta, \qquad \text{with} \qquad \omega \equiv \tfrac{1}{2}(\omega - \omega^*), \quad \beta \equiv \tfrac{1}{2}(\omega + \omega^*). \qquad (2.17)$$

If $\{J_i, i = 1, 2, 3\}$ is a basis for the vector space $\mathfrak{so}(3)$ with:

$$J_1 = \begin{bmatrix} 0 & 0 & 0 & 0 \\ \hline 0 & 0 & 0 & 0 \\ 0 & 0 & 0 & -1 \\ 0 & 0 & 1 & 0 \end{bmatrix}, \qquad J_2 = \begin{bmatrix} 0 & 0 & 0 & 0 \\ \hline 0 & 0 & 0 & 1 \\ 0 & 0 & 0 & 0 \\ 0 & -1 & 0 & 0 \end{bmatrix}, \qquad J_3 = \begin{bmatrix} 0 & 0 & 0 & 0 \\ \hline 0 & 0 & -1 & 0 \\ 0 & 1 & 0 & 0 \\ 0 & 0 & 0 & 0 \end{bmatrix}, \qquad (2.18)$$

and $\{K_i, i = 1, 2, 3\}$ is a basis for $\mathfrak{b}(3)$ with:

$$K_1 = \begin{bmatrix} 0 & 1 & 0 & 0 \\ \hline 1 & 0 & 0 & 0 \\ 0 & 0 & 0 & 0 \\ 0 & 0 & 0 & 0 \end{bmatrix}, \qquad K_2 = \begin{bmatrix} 0 & 0 & 1 & 0 \\ \hline 0 & 0 & 0 & 0 \\ 1 & 0 & 0 & 0 \\ 0 & 0 & 0 & 0 \end{bmatrix}, \qquad K_3 = \begin{bmatrix} 0 & 0 & 0 & 1 \\ \hline 0 & 0 & 0 & 0 \\ 0 & 0 & 0 & 0 \\ 1 & 0 & 0 & 0 \end{bmatrix} \qquad (2.19)$$

then the commutation relations for the basis elements will take the form:

$$[J_i, J_j] = \varepsilon_{ijk} J_k, \qquad [J_i, K_j] = \varepsilon_{ijk} K_k, \qquad [K_i, K_j] = -\varepsilon_{ijk} J_k. \qquad (2.20)$$

If we include the infinitesimal translations, while using the canonical basis $\delta_\mu$ for $\mathbb{R}^4$, then in order to extend the commutation relations (2.20) to the Poincaré Lie algebra $\mathfrak{iso}(1, 3)$ one will need only to add the Lie brackets of the $\delta_\mu$ with each other and the $J_i$ and $K_i$. From (2.16) and direct computation, one finds that:

$$\begin{aligned}
&[\delta_\mu, \delta_\nu] = 0, \\
&[\delta_0, J_i] = 0, \qquad [\delta_i, J_j] = \varepsilon_{ijk}\,\delta_k, \\
&[\delta_0, K_i] = -\delta_i, \; [\delta_i, K_j] = -\delta_{ij}\delta_0.
\end{aligned} \qquad (2.21)$$

Thus, one sees that one has extended the Lie algebra of $\mathfrak{iso}(3)$ to the extra basis elements $\delta_0, K_i, i = 1, 2, 3$ of $\mathfrak{iso}(1, 3)$.

*d. The dual space of the Lie algebra* $\mathfrak{iso}(3, 1)$. The dual space to the Lie algebra $\mathfrak{iso}(3, 1)$, which then consists of all linear functionals on $\mathfrak{iso}(3, 1)$, will be denoted by $\mathfrak{iso}(3, 1)^*$. There is then a natural bilinear pairing $\mathfrak{iso}(3, 1)^* \times \mathfrak{iso}(3, 1) \to \mathbb{R}$, $(\alpha, \mathfrak{a}) \mapsto \alpha(\mathfrak{a})$, and although the two vector spaces are isomorphic, the isomorphism is not canonical. In fact, the choice of isomorphism will essentially be a choice of mechanical constitutive law when the elements of the Lie algebra $\mathfrak{iso}(3, 1)$ relate to the infinitesimal deformations of the kinematical state of a moving/deformed object (i.e., virtual displacements or variations), and the elements of $\mathfrak{iso}(3, 1)^*$ define the "conjugate" dynamic/kinetic state.

Generally, elements of the bilinear pairing of dynamical states and virtual displacements of kinematical states will give the virtual work that is done by the virtual displacement. For instance, if the elements of $\mathfrak{iso}(3, 1)$ are regarded as infinitesimal displacements of an embedded object then the elements in $\mathfrak{iso}(3, 1)^*$ will represent the forces and torques that act upon it, while if the elements of $\mathfrak{iso}(3, 1)$ are regarded as variations of its velocities or displacement gradients then the dual elements in $\mathfrak{iso}(3, 1)^*$ will represent its momenta or stresses, resp.

*e. The bundle of oriented, Lorentzian frames on $\mathfrak{M}^4$.* Just as the group manifold $ISO(3)$ was diffeomorphic to the bundle $SO(E^3)$ of oriented, Euclidian orthogonal frames on $E^3$, so is the group manifold $ISO(1, 3)$ diffeomorphic to the bundle $L_0(\mathfrak{M}^4)$ of oriented, Lorentzian frames on Minkowski space. Indeed, the diffeomorphism comes about analogously by choosing some oriented, Lorentzian frame $(O, \mathbf{e}_\mu)$ to associate with the identity element $(0, \delta_\nu^\mu)$ in $ISO(1, 3)$. One then associates any other element $(a^\mu, L_\nu^\mu)$ of $ISO(1, 3)$ with the frame:

$$(x, \overline{\mathbf{e}}_\nu) = (O, \mathbf{e}_\mu)(a^\mu, L_\nu^\mu) = (O + L_\nu^\mu a^\nu, L_\nu^\mu \mathbf{e}_\mu), \qquad (2.22)$$

which can also be expressed as the system of equations:

$$x = O + L_\nu^\mu a^\nu, \qquad\qquad \overline{\mathbf{e}}_\nu = L_\nu^\mu \mathbf{e}_\mu. \qquad (2.23)$$

This also defines a *right* action of the group $ISO(1, 3)$ on the bundle $L_0(\mathfrak{M}^4)$ that, as before in Part I, is not the same as the right action of $ISO(1, 3)$ as a structure group on the bundle $L_0(\mathfrak{M}^4)$ of Poincaré frames on the affine tangent space to $\mathfrak{M}^4$. Once again, the action of any structure group on a principal fiber bundle is vertical – i.e., it projects to the identity in the base manifold – whereas the semi-direct product action that we are using involves the translation of frames from one point to another.

As a bundle, $L_0(\mathfrak{M}^4)$ is trivial, and one can express it as the product manifold $\mathfrak{M}^4 \times SO(1, 3)$, which again amounts to the semi-direct product $\mathbb{R}^4 \times_s SO(1, 3)$ that is associated with $ISO(1, 3)$. Thus, one can also express any of the tangent spaces to the manifold $L_0(\mathfrak{M}^4)$ – i.e., the total space to the bundle – as the direct sum $\mathbb{R}^4 \oplus \mathfrak{so}(1, 3)$, which is the vector space that underlies the Lie algebra $\mathfrak{iso}(1, 3)$. Thus, a typical tangent vector to $L_0(\mathfrak{M}^4)$ can be expressed in the component form:

$$\mathbf{X} = \boldsymbol{\xi} + \boldsymbol{\eta} = X^\mu \frac{\partial}{\partial x^\mu} + X_\mu \frac{\partial}{\partial \mathbf{e}_\mu}, \tag{2.24}$$

or, after one embeds the bundle in the manifold $\mathbb{R}^4 \times M(4)$, where $M(4)$ refers to the vector space of all 4×4 real matrices in the form:

$$\mathbf{X} = \tilde{X}^\mu \frac{\partial}{\partial a^\mu} + \tilde{X}^\mu_\nu \frac{\partial}{\partial I^\mu_\nu}. \tag{2.25}$$

Since the action of $ISO(1, 3)$ on the bundle $L_0(\mathfrak{M}^4)$ amounts to the right action of $ISO(1, 3)$ on itself when one identifies it with the aforementioned bundle, there is a corresponding right action of the Lie algebra $\mathfrak{iso}(1, 3)$ on the tangent spaces to $L_0(\mathfrak{M}^4)$ that amounts to the right action of $\mathfrak{iso}(1, 3)$ on itself when one identifies the underlying vector space to $\mathfrak{iso}(1, 3)$ with each tangent space. Thus, if $\mathbf{X}$ has the form (2.24) and the element of $\mathfrak{iso}(1, 3)$ is $\boldsymbol{\Omega} = \mathbf{v} + \boldsymbol{\omega}$ then the right action of $\boldsymbol{\Omega}$ on $\mathbf{X}$ will be:

$$[\mathbf{X}, \boldsymbol{\Omega}] = [\mathbf{x} + \boldsymbol{\eta}, \mathbf{v} + \boldsymbol{\omega}] = \boldsymbol{\eta}\mathbf{v} - \boldsymbol{\omega}\mathbf{x} + [\boldsymbol{\eta}, \boldsymbol{\omega}]. \tag{2.26}$$

Thus, the translational part of the result is:

$$\boldsymbol{\eta}\mathbf{v} - \boldsymbol{\omega}\mathbf{x} = (\eta^\mu_\nu v^\nu - \omega^\mu_\nu x^\nu)\frac{\partial}{\partial a^\mu} \tag{2.27}$$

and the Lorentzian part is:

$$[\boldsymbol{\eta}, \boldsymbol{\omega}] = (\eta^\mu_\kappa \omega^\kappa_\nu - \omega^\mu_\kappa \eta^\kappa_\nu)\frac{\partial}{\partial I^\mu_\nu}; \tag{2.28}$$

that is, (2.27) is the part that is tangent to $\mathbb{R}^4$ and (2.28) is the part that is tangent to $SO(1, 3)$, respectively.

One can also locally represent the fundamental vector field $\tilde{\mathfrak{a}}(x, \mathbf{e}_\mu)$ on $L_0(\mathfrak{M}^4)$ that is associated with $\mathfrak{a} = (v^\mu, \omega^\mu_\nu) \in \mathfrak{iso}(3, 1)$ as:

$$\tilde{\mathfrak{a}}(x, \mathbf{e}_\mu) = v^\mu \frac{\partial}{\partial x^\mu} + \omega^\mu_\nu \frac{\partial}{\partial I^\mu_\nu}, \tag{2.29}$$

in which we have essentially associated the frame $(\boldsymbol{\delta}_\mu, I^\mu_\nu)$ for $\mathfrak{iso}(3, 1)$ with the local frame field $\left(\frac{\partial}{\partial x^\mu}, \frac{\partial}{\partial I^\mu_\nu}\right)$ on some open subset of $L_0(\mathfrak{M}^4)$.

**3. Kinematics of relativistic Cosserat media.** Since the essential character of the discussion in Part I of the mechanics of objects non-relativistic Cosserat media had more to do with the fact that we were starting with a group that was a semi-direct product of translations and rotations than the fact that the rotations were Euclidian, we find that all that is required at this point is to specialize a few general notions to the Poincaré group by discussing the role of the extra dimensions in it beyond those of its $ISO(3)$ subgroups.

*a. Kinematical state of a relativistic Cosserat object.* What we are calling a *relativistic Cosserat medium* is any manifold that takes the form of $L_0(\mathfrak{M}^4)$, and an *object* in that medium will then take the form of an embedded submanifold $f \colon \mathcal{O} \to u^a \mapsto (x(u), \mathbf{e}_\mu(u))$, where $\mathcal{O} \subset \mathbb{R}^p$ is a subset of the space of material coordinates for the object being embedded. One can then think of $f$ as consisting of an object $\tilde{f} \colon \mathcal{O} \to \mathfrak{M}^4$, $u^a \mapsto x(u)$ in Minkowski space and a Lorentzian frame field $\mathbf{e} \colon \mathcal{O} \to L_0(\mathfrak{M}^4)$, $u^a \mapsto \mathbf{e}_\mu(u)$ on $\tilde{f}$.

Note that this definition is more general than the association of the frame $\mathbf{e}_\mu(u)$ with $x(u)$ directly, which amounts to defining a Lorentzian frame field on the image $\tilde{f}(\mathcal{O})$ of $\mathcal{O}$ in $\mathfrak{M}^4$, since frame fields do not always have to exist on that image globally, but only locally. Thus, we are really defining frame fields on $\tilde{f}(\mathcal{O})$ that are potentially singular. Furthermore, the number of frame elements – viz., four – does not have to equal the dimension of the object; for instance, that will be the case when one defines a Lorentzian frame field along a curve.

The *kinematical state* of the object $\mathcal{O}$ is a section $s \colon \mathcal{O} \to J^1(\mathcal{O}; L_0(\mathfrak{M}^4))$ of the source projection $J^1(\mathcal{O}; L(\mathfrak{M}^4)) \to \mathcal{O}$, $j_u^1 f \mapsto u$ of the manifold $J^1(\mathcal{O}; L_0(\mathfrak{M}^4))$ of 1-jets of local embeddings of $\mathcal{O}$ into $L_0(\mathfrak{M}^4)$. Since a local coordinate system for $J^1(\mathcal{O}; L_0(\mathfrak{M}^4))$ looks like $(u^a, x^\mu, e_\nu^\mu, x_a^\mu, e_{\nu a}^\mu)$, such a section will take the local form:

$$s(u) = (u^a, x^\mu(u), e_\nu^\mu(u), x_a^\mu(u), e_{\nu a}^\mu(u)). \tag{2.30}$$

The section $s(u)$ is *integrable* iff it is the 1-jet prolongation $j^1 f$ of some object $f$ in $L_0(\mathfrak{M}^4)$, in which case one has the local conditions:

$$x_a^\mu(u) = \frac{\partial x^\mu}{\partial u^a}, \qquad e_{\nu a}^\mu(u) = \frac{\partial e_\nu^\mu}{\partial u^a}. \tag{2.31}$$

One then desires to define the "deformation" that takes an initial state $s_0$ to some final state $s$. One way of doing this is to define a deformation of the initial object $f_0$ into the final object $f$ and then prolong it – i.e., differentiate it – to a corresponding map of sections. One can then generalize to non-integrable sections by essentially "dropping the commas" in the derivative coordinates and making the transformation purely algebraic.

*b. The action of* $J^1(\mathcal{O}; ISO(3, 1))$ *on* $J^1(\mathcal{O}; L_0(\mathfrak{M}^4))$. The way that we effected the deformation of kinematical states in Part I was to let $J^1(\mathcal{O}; ISO(3))$ act upon the manifold $J^1(\mathcal{O}; SO(E^3))$, where we generally took that to mean the action of the group of sections of one source projection on sections of the other. Here, we shall simply replace the space $E^3$ and group $ISO(3)$ with $\mathfrak{M}^4$ and $ISO(3, 1)$, respectively.

One starts with the left action of $C^\infty(\mathcal{O}; ISO(3))$ on the components of $C^\infty(\mathcal{O}; L_0(\mathfrak{M}^4))$:

$$x^\mu(\rho) = a^\mu(\rho) + L^\mu_\nu(\rho)x^\nu_0(\rho), \qquad e^\mu_\nu(\rho) = e^\mu_{0\kappa}(\rho)L^\kappa_\nu(\rho), \qquad (2.32)$$

and differentiate with respect to $\rho^a$, while dropping the explicit dependence upon $\rho$ and the commas:

$$x^\mu_a = a^\mu_a + L^\mu_{\nu a}x^\nu_0 + L^\mu_\nu x^\nu_{0a} \qquad e^\mu_{\nu a} = L^\mu_{\kappa a}e^\kappa_{0\nu} + L^\mu_\kappa e^\kappa_{0\nu a}. \qquad (2.33)$$

Equations (2.32) and (2.33) collectively specify how the section $(\rho^a, a^\mu(\rho), L^\mu_\nu(\rho), a^\mu_a(\rho), L^\mu_{\nu a}(\rho))$ acts on the initial kinematical state $(\rho^a, x^\mu_0(\rho), e^\mu_{0\nu}(\rho), x^\mu_{0a}(\rho), e^\mu_{0\nu a}(\rho))$ to produce the deformed state $(\rho^a, x^\mu(\rho), e^\mu_\nu(\rho), x^\mu_a(\rho), e^\mu_{\nu a}(\rho))$. We shall then think of the section $\chi(\rho) = (\rho^a, a^\mu(\rho), L^\mu_\nu(\rho), a^\mu_a(\rho), L^\mu_{\nu a}(\rho))$ as the *displacement* that takes the initial state to the deformed state by way of the action:

$$s(\rho) = \chi(\rho) \cdot s_0(\rho). \qquad (2.34)$$

It is also important to see how the deformation will appear from a co-moving or co-deformed frame field, such as $(x, \mathbf{e}_\mu)$. One must then replace $x^\mu_0$ and $e^\mu_{0\nu}$ in (2.33) with the results of solving (2.32) for those two sets of coordinates:

$$x^\mu_0 = \tilde{L}^\mu_\nu(x^\nu - a^\nu), \qquad e^\mu_{0\nu} = \tilde{L}^\mu_\kappa e^\kappa_\nu. \qquad (2.35)$$

This gives:

$$x^\mu_a = \xi^\mu_a + \omega^\mu_{\nu a}x^\nu + L^\mu_\nu x^\nu_{0a} \qquad e^\mu_{\nu a} = L^\mu_\kappa e^\kappa_{0\kappa a} + \omega^\mu_{\kappa a}e^\kappa_\nu, \qquad (2.36)$$

into which we have introduced:

$$\xi^\mu_a = a^\mu_a - \omega^\mu_{\nu a}a^\nu, \qquad \omega^\mu_{\nu a} = L^\mu_{\kappa a}\tilde{L}^\kappa_\nu. \qquad (2.37)$$

We then see that $(\xi^\mu_a, \omega^\mu_{\nu a})$, which belongs to $C^\infty(\mathcal{O}; \mathfrak{iso}(3))$, can be expressed in the form:

$$(\xi^\mu_a, \omega^\mu_{\nu a}) = (a^\mu_a, L^\mu_{\nu a})(-\tilde{L}^\mu_\nu a^\nu, \tilde{L}^\mu_\nu) = dg \cdot g^{-1}, \qquad (2.38)$$

where $g = (a^\mu, L^\mu_\nu)$.

Hence, the displacement $\chi(\rho)$ can also be expressed in the form:

$$\bar{\chi}(\rho) = (\rho^{\mu}, 0, \ \delta^{\mu}_{\nu}, \ \xi^{\mu}_{a}(\rho), \ \omega^{\mu}_{\nu a}(\rho)) = \chi(\rho) \cdot g^{-1}. \tag{2.39}$$

We then introduce the notation:

$$\nabla g = dg \cdot g^{-1}, \qquad \nabla a^{\mu} = da^{\mu} - \omega^{\mu}_{\nu} a^{\nu}, \qquad \nabla L^{\mu}_{\nu} = dL^{\mu}_{\kappa} \tilde{L}^{\kappa}_{\nu}, \tag{2.40}$$

so we also have defined a 1-form $(\xi^{\mu}, \omega^{\mu}_{\nu})$ on $\mathcal{O}$ with values in $\mathfrak{iso}(3, 1)$:

$$\xi^{\mu} = \nabla a^{\mu}, \qquad \omega^{\mu}_{\nu} = \nabla L^{\mu}_{\nu}. \tag{2.41}$$

We then see that $\nabla$ defines a map $\nabla \colon \Lambda^0(\mathcal{O}) \otimes ISO(1, 3) \to \Lambda^1(\mathcal{O}) \otimes \mathfrak{iso}(1, 3)$, $(a, L) \mapsto (\nabla a, \nabla L)$. More concisely, one can say that it takes $g$ to $\nabla g$.

We will regard either $dg$ or $\nabla g$ as the *deformation* of the displacement $g$. The former refers to the Lagrangian picture, while the latter refers to the Eulerian picture. We use the notations for deformations:

$$E(\rho) = (da^{\mu}, \ dL^{\mu}_{\nu}) \text{ or } (\xi^{\mu}, \omega^{\mu}_{\nu}), \tag{2.42}$$

depending upon which picture we are using.

It is important to see that, just as the action of $ISO(1, 3)$ on $L_0(\mathfrak{M}^4)$ was not the action of a structure group on a principal fiber bundle, neither is the 1-form $\nabla g$ the 1-form of a connection on $L_0(\mathfrak{M}^4)$ with values in $\mathfrak{so}(1, 3)$, in the Cartan sense of connections on frame bundles. In particular, the 1-forms $\nabla a^{\mu}$ do not have to represent a coframe on $\mathbb{R}^4$, and might even be zero.

  *c. Integrability of deformations.* One can now address the issue of the compatibility – or *integrability* – of a deformation $E$; that is, the integrability of either system of differential equations:

$$E = dg \qquad \text{or} \qquad E = \nabla g, \tag{2.43}$$

depending upon whether one is using the Lagrangian or Eulerian picture, respectively.

When one is describing a deformation in the Lagrangian picture, this is simply a matter of the exactness of a differential 1-form. Namely, if $E = (\upsilon^{\mu}, \varpi^{\mu}_{\nu})$ is supposed to represent an exact 1-form on $\mathcal{O}$ with values in $T(ISO(3, 1))$ – viz., $E = dg = (da^{\mu}, \ dL^{\mu}_{\nu})$ – then one must have that $E$ is closed:

$$d_{\wedge} E = (0, 0); \tag{2.44}$$

i.e.:

$$d_{\wedge} \upsilon^{\mu} = 0, \qquad d_{\wedge} \varpi^{\mu}_{\nu} = 0. \tag{2.45}$$

Whether this is also sufficient depends upon whether $\mathcal{O}$ is also simply-connected. At the moment, we are only considering a vector space, so this is not an issue, but since dislocations and disclinations represent non-trivial generators to the fundamental group of $\mathcal{O}$, one should not abandon all such considerations out of hand; however, we will defer that discussion to a later study.

On the other hand, when the deformation $E$ is regarded in the Eulerian picture, so $E = (\xi^{\mu}, \omega^{\mu}_{\nu}) \in \Lambda^1(\mathcal{O}) \otimes \mathfrak{iso}(1, 3)$, one will find, by taking the exterior derivatives of both sets of equations in (2.41) that:

$$d_{\wedge} \xi = \omega \,^{\wedge} \xi, \qquad\qquad d_{\wedge} \omega = \omega \,^{\wedge} \omega. \qquad\qquad (2.46)$$

Thus, if we extend the previous $\nabla$ to a linear operator $\nabla \colon \Lambda^1(\mathcal{O}) \otimes \mathfrak{iso}(1, 3) \to \Lambda^2(\mathcal{O}) \otimes \mathfrak{iso}(1, 3)$, $(\xi, \omega) \mapsto (\nabla \xi, \nabla \omega)$, with:

$$\nabla \xi = d_{\wedge} \xi - \omega \,^{\wedge} \xi, \qquad \nabla \omega = d_{\wedge} \omega - \omega \,^{\wedge} \omega, \qquad\qquad (2.47)$$

then we will see that the necessary condition for the deformation $E$ to be integrable into $\nabla g$ for some displacement $g$ is that:

$$\nabla E = 0, \qquad\qquad (2.48)$$

which is to say, (2.46).

Thus, for the two successive applications of the operator $\nabla$, as we have defined it, one has:

$$\nabla^2 = 0. \qquad\qquad (2.49)$$

*d. Dislocations.* What we will call the *dislocation* ([1]) in a deformation $(\xi^{\mu}, \omega^{\mu}_{\nu})$ is then the 2-form $\Omega = \nabla E = (\Omega^{\mu}, \Omega^{\mu}_{\nu})$ with values in $\mathfrak{iso}(1, 3)$, in which:

$$\Omega^{\mu} = d_{\wedge} \xi^{\mu} - \omega^{\mu}_{\nu} \,^{\wedge} \xi^{\nu}, \qquad\qquad \Omega^{\mu}_{\nu} = d_{\wedge} \omega^{\mu}_{\nu} - \omega^{\mu}_{\kappa} \wedge \omega^{\kappa}_{\nu}. \qquad\qquad (2.50)$$

Thus, the necessary condition for the compatibility of $E$ – i.e., its integrability into $\nabla g$ – is that $\Omega$ must vanish. In the contrary case, there would be no displacement that could produce the deformation $E$.

Just as we have been emphasizing that the action of the Poincaré group on Lorentzian frames is not the action of a structure group on a principal fiber bundle, and that $\omega$ is not really a connection 1-form, similarly, we point out that despite the superficial resemblance between (2.50) and the Cartan structure equations for torsion and curvature, that identification would not be precise, either.

It is also important to note that dislocations can also have purely topological origins, and, in fact, that was what Volterra, and later Armanni [3], were describing. For Volterra, dislocations involved deformations $\omega$ for which $E$ vanished, although they still did not admit any displacement $g$; that is, one had $\nabla \omega = 0$, but not $\omega = \nabla g$. The root of that situation was in the multiple connectivity of the object $\mathcal{O}$, which meant that the one-dimensional de Rham cohomology vector space $H^1_{dR}(\mathcal{O})$ had a positive dimension. However, what Armanni then pointed out in a note that was communicated by Volterra was that dislocations could come from non-trivial two-dimensional de Rham cohomology, as well.

---

([1]) Actually, since dislocations are associated with infinitesimal translations and disclinations are associated with infinitesimal rotations, to call $\Omega$ the "dislocation" is an abbreviation, since it really includes both. Volterra [2] originally called both of them collectively "distortions," and some refer to disclinations as simply "rotational dislocations."

The use of the word "dislocation" to refer to the 2-form $\Omega$ is more related to the later work in the theory, which passed from individual dislocation lines as the sources of self-stresses to continuous distributions of dislocations that were characterized by the present 2-form, in effect. Of course, there is a crucial difference between dislocations that come from the non-vanishing of a de Rham cohomology space, while $\Omega$ still vanishes, and those that are purely due to the non-vanishing of $\Omega$. In those aforementioned later works, the equation $\nabla E = \Omega$ was, in effect, referred to as the "fundamental geometric equation of dislocation theory;" the 2-form $\Omega$ then served as the source of the deformation $E$.

We can continue the pattern that we have established up to now by looking for the necessary, but not always sufficient, condition for a given 2-form $\Omega = (\Omega^\mu, \ \Omega^\mu_\nu)$ to be the dislocation of some deformation $E$; that is $\Omega = \nabla E$. Basically, one takes the exterior derivatives of both sets of 2-forms in $E$, using the definition (2.50), and gets:

$$d_\nabla \Omega^\mu = -\ \Omega^\mu_\nu \wedge \xi^\nu + \omega^\mu_\nu \wedge \Omega^\mu, \qquad d_\wedge \Omega^\mu_\nu = \omega^\mu_\kappa \wedge \Omega^\kappa_\nu - \Omega^\mu_\kappa \wedge \omega^\kappa_\nu. \qquad (2.51)$$

This is then the necessary, but not sufficient, condition for the 2-form $\Omega$ to represent the source of a deformation $E$.

If we then define the *incompatibility* 3-form $\Psi = (\Psi^\mu, \ \Psi^\mu_\nu) = \nabla \Omega$ with:

$$\Psi^\mu = \ d_\nabla \Omega^\mu \ + \Omega^\mu_\nu \wedge \ \xi^\nu - \omega^\mu_\nu \wedge \Omega^\mu, \quad \Psi^\mu_\nu = d_\wedge \Omega^\mu_\nu + \Omega^\mu_\kappa \wedge \omega^\kappa_\nu - \omega^\mu_\kappa \wedge \Omega^\kappa_\nu \qquad (2.52)$$

then we can rephrase this condition as $\Psi = 0$. Hence, in the contrary case, $\Omega$ could not be the dislocation of any deformation.

As before, although (2.51) are formally similar to the Bianchi identities for the torsion and curvature of a connection, in the present case, there is a subtle distinction that is due to the fact that we are not starting with a connection. Furthermore, it is important to notice that while the compatibility conditions for $E$ could be expressed in terms of only its component 1-forms $v^\mu$ and $\omega^\mu_\nu$, the compatibility conditions for $\Omega$ involve not only the component 2-forms $\Omega^\mu$ and $\Omega^\mu_\nu$, but also the component 1-forms $v^\mu$ and $\omega^\mu_\nu$. That is, one must choose a deformation $E$, even though the non-vanishing of $\Psi$ would imply its non-existence, at least in the sense that $\Omega$ would represent $\nabla E$.

Now that we have defined all of its terms, we can summarize the kinematical definitions into following the sequence of (infinite-dimensional) vector spaces and linear maps (after the first term):

$$ISO(1, 3) \longrightarrow \Lambda^0 \otimes ISO(1, 3) \ \xrightarrow{\ \nabla\ } \ \Lambda^1 \otimes \mathfrak{iso}(1, 3) \ \xrightarrow{\ \nabla\ } \ \Lambda^2 \otimes \mathfrak{iso}(1, 3) \ \xrightarrow{\ \nabla\ } \ \Lambda^3 \otimes \mathfrak{iso}(1, 3)$$

The first arrow refers to the inclusion of the Poincaré group in the constant functions on $\mathcal{O}$ with values in $ISO(1,3)$. We have also abbreviated $\Lambda^k(\mathcal{O})$ to simply $\Lambda^k$, in each case. Although this sequence could be continued to the right, that will not be necessary in what follows. Indeed, the key step in the eyes of statics/dynamics is defined by the second arrow.

Although this latter sequence is cohomological, as a result of (2.49), it does not have to be exact. That is, although the image of $\nabla$ from one step is a linear subspace of the kernel of $\nabla$ at the next step, the two linear spaces do not have to be identical. The quotient vector space is then the $\nabla$-cohomology vector space in that dimension. However, such investigations go beyond the scope of the present study, so we defer them to a later examination.

**4. Statics/dynamics in relativistic Cosserat media.** Since we have already introduced the basic picture for defining the dynamical state of a Cosserat medium and its equations of equilibrium or time evolution in Part I, in the context of non-relativistic motion, all that we really need to do at this point is to adapt the definitions that were made for three-dimensional Euclidian rigid motions to four-dimensional Minkowskian "rigid" motions.

*a. Virtual displacements of the kinematical state.* If $\psi(\rho)$ is a relativistic Cosserat object then a *virtual displacement* $\delta\psi$ of that object will be defined to be a vector field on the image of $\mathcal{O}$ in $L_0(\mathfrak{M}^4)$. It will then have the local form:

$$\delta\psi(\rho) = \delta x^\mu(\rho)\frac{\partial}{\partial x^\mu} + \delta e_\nu^\mu(\rho)\frac{\partial}{\partial e_\nu^\mu}. \tag{3.1}$$

A virtual displacement $\delta s(\rho)$ of a kinematical state $s(\rho)$ is then a vector field on the image of $\mathcal{O}$ in $J^1(\mathcal{O}; L_0(\mathfrak{M}^4))$. In the natural frame it will have the local form:

$$\delta s(\rho) = \delta x^\mu(\rho)\frac{\partial}{\partial x^\mu} + \delta e_\nu^\mu(\rho)\frac{\partial}{\partial e_\nu^\mu} + \delta x_b^\mu(\rho)\frac{\partial}{\partial x_b^\mu} + \delta e_{\nu b}^\mu(\rho)\frac{\partial}{\partial e_{\nu b}^\mu} \tag{3.2}$$

in the Lagrangian picture.

In the Eulerian picture, it will take the form:

$$\delta s(\rho) = \delta\xi^\mu \boldsymbol{\xi}_\mu + \delta I_\nu^\mu \boldsymbol{\mathcal{I}}_\mu^\nu + \delta\xi_a^\mu \boldsymbol{\xi}_\mu^a + \delta I_{\nu a}^\mu \boldsymbol{\mathcal{I}}_\mu^{\nu a}, \tag{3.3}$$

in which the anholonomic frame field $\boldsymbol{\xi}_\mu$, $\boldsymbol{\mathcal{I}}_\mu^\nu$, $\boldsymbol{\xi}_\mu^a$, $\boldsymbol{\mathcal{I}}_\mu^{\nu a}$ can be expressed in terms of the natural frame field, although the specific expressions will not be needed later in this study.

Just as the deformed state could be expressed as the action of $J^1(\mathcal{O}; ISO(3, 1))$ on initial state in $J^1(\mathcal{O}; L_0(\mathfrak{M}^4))$, similarly, the virtual displacement of a kinematical state can be described as the action of $J^1(\mathcal{O}; \mathfrak{iso}(3, 1))$ on $J^1(\mathcal{O}; L_0(\mathfrak{M}^4))$ to produce a vector field on $J^1(\mathcal{O}; L_0(\mathfrak{M}^4))$, namely, the fundamental vector field of some section of the source projection of $J^1(\mathcal{O}; \mathfrak{iso}(3, 1))$. The way that one gets the local equations for this action is by applying $\delta$ to the equations (2.32) and (2.33) of the group action as a linear differential operator:

$$\delta x^{\mu} = \delta a^{\mu} + \delta L^{\mu}_{\nu} x^{\nu}_0 \,, \qquad\qquad \delta e^{\mu}_{\nu} = \delta L^{\mu}_{\kappa} e^{\kappa}_{0\nu} \,, \tag{3.4}$$

$$\delta x^{\mu}_a = \delta a^{\mu} + \delta L^{\mu}_{\nu a} x^{\nu}_0 + \delta L^{\mu}_{\nu} x^{\nu}_{0a} \,, \qquad\qquad \delta e^{\mu}_{\nu a} = \delta L^{\mu}_{\kappa a} e^{\kappa}_{0\nu} + \delta L^{\mu}_{\kappa} e^{\kappa}_{0\nu a} \,. \tag{3.5}$$

(Although we should really be including variations of the initial state, since that is not typical of the calculus of variations, we shall omit them.)

In order to obtain the Cosserat equations, we shall require the form that this takes in the Eulerian picture. By solving the group action for $x^{\nu}_0$ and $e^{\kappa}_{0\nu}$, the latter equations will take the form:

$$\delta x^{\mu} = \delta \xi^{\mu} + \delta I^{\mu}_{\nu} x^{\nu} \,, \qquad\qquad \delta e^{\mu}_{\nu} = \delta I^{\mu}_{\kappa} e^{\kappa}_{\nu} \,, \tag{3.6}$$

$$\delta x^{\mu}_a = \delta \xi^{\mu}_a + \delta I^{\mu}_{\nu a} x^{\nu} + \delta I^{\mu}_{\nu} x^{\nu}_a \,, \qquad\qquad \delta e^{\mu}_{\nu a} = \delta I^{\mu}_{\kappa a} e^{\kappa}_{\nu} + \delta I^{\mu}_{\kappa} e^{\kappa}_{\nu a} \,. \tag{3.7}$$

into which we have introduced:

$$\delta \xi^{\mu} = \delta a^{\mu} - \delta I^{\mu}_{\nu} a^{\nu} \,, \qquad\qquad \delta I^{\mu}_{\nu} = \delta L^{\mu}_{\kappa} \tilde{L}^{\kappa}_{\nu} \,, \tag{3.8}$$

and differentiated (3.6) to get (3.7). That will then make:

$$\delta \xi^{\mu}_a = \delta a^{\mu}_a - \delta I^{\mu}_{\nu a} a^{\nu} - \delta I^{\mu}_{\nu} a^{\nu}_a \,, \qquad\qquad \delta I^{\mu}_{\nu a} = \delta L^{\mu}_{\kappa a} \tilde{L}^{\kappa}_{\nu} + \delta L^{\mu}_{\kappa} \tilde{L}^{\kappa}_{\nu a} \,. \tag{3.9}$$

Thus, we now have the action of a section $\chi(\rho) = (\rho^i, \, \delta \xi^{\mu}(r), \, \delta I^{\mu}_{\nu}(\rho), \, \delta \xi^{\mu}_a(\rho), \, \delta I^{\mu}_{\nu a}(\rho))$ of the source projection of $J^1(\mathcal{O}; \mathfrak{iso}(3, 1))$ on a section $s(\rho) = (\rho^i, \, x^{\mu}(\rho), \, e^{\mu}_{\nu}(\rho), \, x^{\mu}_a(\rho), \, e^{\mu}_{\nu a}(\rho))$ to produce the vector field $\delta s(\rho)$ on $s(\rho)$, which then represents the fundamental vector field that is associated with $\chi(\rho)$ by way of the action of $J^1(\mathcal{O}; ISO(3, 1))$ on the jet manifold $J^1(\mathcal{O}; L_0(\mathfrak{M}^4))$.

*b. Integrable variations of kinematical states.* If the virtual displacement $\delta s$ is *integrable* then, by definition, it will be the 1-jet prolongation of a vector field $\delta \psi$ on the object in $L_0(\mathfrak{M}^4)$:

$$\delta s = \delta^1 \psi \equiv j^1 \delta \psi \,; \tag{3.10}$$

$\delta s$ will then take the local form:

$$j^1 \delta \psi = \delta x^\mu(\rho) \frac{\partial}{\partial x^\mu} + \delta e_\nu^\mu(\rho) \frac{\partial}{\partial e_\nu^\mu} + \frac{\partial(\delta x^\mu)}{\partial \rho^b}(\rho) \frac{\partial}{\partial x_b^\mu} + \frac{\partial(\delta e_\nu^\mu)}{\partial \rho^b}(\rho) \frac{\partial}{\partial e_{\nu b}^\mu} \qquad (3.11)$$

in the Lagrangian picture; i.e.:

$$\delta x_a^\mu = \frac{\partial(\delta x^\mu)}{\partial \rho^a}, \qquad \delta e_{\nu a}^\mu = \frac{\partial(\delta e_\nu^\mu)}{\partial \rho^a}. \qquad (3.12)$$

As for the Eulerian picture, in order to make (3.9) consistent with this, one must then have:

$$\delta \xi_a^\mu = \frac{\partial(\delta \xi^\mu)}{\partial \rho^a}, \qquad \delta I_{\nu a}^\mu = \frac{\partial(\delta I_\nu^\mu)}{\partial \rho^a}, \qquad x_a^\mu = \frac{\partial x^\mu}{\partial \rho^a}, \qquad e_{\nu a}^\mu = \frac{\partial e_\nu^\mu}{\partial \rho^a}. \qquad (3.13)$$

Thus, both the sections $\chi(\rho)$ and $s(\rho)$ must be integrable.

*c. The dynamical state of an object.* We define the dynamical state of a relativistic Cosserat object by a 1-form on the object that we call the *fundamental 1-form*. It is dual to the virtual displacement of the object under the bilinear pairing of 1-forms and vector fields that gives the density of virtual work, and thus expresses the response of the object to the virtual displacement. This 1-form locally looks like:

$$\phi = F_\mu \, dx^\mu + M_\mu^\nu de_\nu^\mu + \sigma_\mu^a dx_a^\mu + \mu_\mu^{\nu a} de_{\nu a}^\mu \qquad (3.14)$$

in the Lagrangian picture.

In this expression:

1. The components $F_\mu$ represent the external body force resultant that is applied to $\mathcal{O}$.

2. The components $M_\mu^\nu$ represent the resultant external body force-couple (i.e., torque) that is applied to $\mathcal{O}$.

3. The components $\sigma_\mu^a$ represent the energy-momentum-stress state of $\mathcal{O}$.

4. The components $\mu_\mu^{\nu a}$ represent the angular-momentum-couple-stress state of $\mathcal{O}$.

The specific functional forms of the components of $\phi$ then embody the mechanical constitutive laws for the medium $\mathfrak{M}^4$ and the object $\mathcal{O}$. In particular, $F_\mu$ might depend upon the point in Minkowski space or the generalized velocity, as in the case of viscosity.

If one differentiates the action of $J^1(\mathcal{O}; ISO(3, 1))$ on $J^1(\mathcal{O}; L_0(\mathfrak{M}^4))$ then one will get:

$$dx^\mu = \xi^\mu + \mathcal{I}_\nu^\mu x^\nu + L_\nu^\mu dx_0^\nu, \qquad (3.15)$$

$$de_\nu^\mu = \mathcal{I}_\kappa^\mu e_\nu^\kappa + L_\kappa^\mu de_{0\nu}^\kappa \,, \tag{3.16}$$

$$dx_a^\mu = \xi_a^\mu + \mathcal{I}_{\nu a}^\mu x^\nu + \mathcal{I}_\nu^\mu x_a^\nu + L_{\nu a}^\mu dx_0^\nu + L_\nu^\mu dx_{0a}^\nu \,, \tag{3.17}$$

$$de_{\nu a}^\mu = \mathcal{I}_{\kappa a}^\mu e_\nu^\kappa + \mathcal{I}_\kappa^\mu e_{\nu a}^\kappa + L_{\kappa a}^\mu de_{0\nu}^\kappa + L_\kappa^\mu de_{0\nu a}^\kappa \,, \tag{3.18}$$

into which we have introduced the anholonomic coframe field:

$$\xi^\mu = da^\mu - \mathcal{I}_\nu^\mu a^\nu \,, \qquad\qquad \mathcal{I}_\nu^\mu = dL_\kappa^\mu \, \tilde{L}_\nu^\kappa \,, \tag{3.19}$$

$$\xi_a^\mu = da_a^\mu - \mathcal{I}_{\nu a}^\mu a^\nu - \mathcal{I}_\nu^\mu a_a^\nu \,, \qquad\qquad \mathcal{I}_{\nu a}^\mu = (dL_{\kappa a}^\mu - \mathcal{I}_\lambda^\mu \, L_{\nu a}^\lambda) \tilde{L}_\nu^\kappa \,. \tag{3.20}$$

With these expressions, one can decompose $\phi$ into a sum $\phi_g + \phi_0$, where $\phi_g$ is a 1-form on $J^1(\mathcal{O}; ISO(3, 1))$ and $\phi_0$ is a 1-form on the initial state. Since we shall have no use for the latter expression, we shall only discuss the former one.

By substituting (3.15) through (3.18) into (3.14), one finds that $\phi_g$ takes the form:

$$\phi_g = F_\mu \, \xi^\mu + \bar{M}_{\mu\nu} \mathcal{I}^{\mu\nu} + \sigma_\mu^a \xi_a^\mu + \bar{\mu}_{\mu\nu}^a \mathcal{I}_a^{\mu\nu} \,, \tag{3.21}$$

in which we have defined:

$$\bar{M}_\mu^{\ \nu} = M_{[\mu\nu]} + F_{[\mu} x_{\nu]} + \sigma_{[\mu|\kappa}^a x_{\nu]a} + \mu_{[\mu|\kappa}^a e_{\nu]a}^\kappa \,, \qquad\qquad \bar{\mu}_{\mu\nu}^a = \mu_{[\mu|\kappa}^a e_{\nu]}^\kappa + \sigma_{[\mu}^a x_{\nu]} \,. \tag{3.22}$$

The anti-symmetrization of the components comes about because of the anti-symmetry in the matrices $\mathcal{I}^{\mu\nu}$ and $\mathcal{I}_a^{\mu\nu}$ in $\mu$ and $\nu$.

If the 1-form $\phi$ is exact then there will exist some function $\mathcal{L}$ on $J^1(\mathcal{O}; L(\mathfrak{M}^4))$ such that $\phi = d\mathcal{L}$. In such a case:

$$F_\mu = \frac{\partial \mathcal{L}}{\partial x^\mu} \,, \qquad M_\mu^{\ \nu} = \frac{\partial \mathcal{L}}{\partial e_\nu^\mu} \,, \qquad \sigma_\mu^a = \frac{\partial \mathcal{L}}{\partial x_a^\mu} \,, \qquad \mu_\mu^{\nu a} = \frac{\partial \mathcal{L}}{\partial e_{\nu a}^\mu} \,. \tag{3.23}$$

The 1-form $\phi$ is then related to the first variation of the action functional that is defined by the Lagrangian density $\mathcal{L}$. However, as we shall see shortly, one can still pose a variational problem using the fundamental 1-form, even when it is not exact, such as when one considers non-conservative forces and non-holonomic constraints.

*d. Virtual work.* The *density of virtual work* that is done by the virtual displacement $\delta s$ of the kinematical state $s$ takes the form:

$$\delta W = \phi(\delta s). \tag{3.24}$$

Thus, when one evaluates $\phi$ on $\delta s$, the result will be a function on the image of $s$ that takes the form:

$$\phi(\delta s) = F_\mu \; \delta x^\mu + M_\mu^{\;\nu} \delta e_\nu^\mu + \sigma_\mu^b \delta x_b^\mu + \mu_\mu^{\nu b} \delta e_{\nu b}^\mu \qquad (3.25)$$

in the Lagrangian picture.

One finds that this, in turn, reverts to:

$$\phi(\delta s) = F_\mu \delta \xi^\mu + \bar{M}_{\mu\nu} \delta I^{\mu\nu} + \sigma_\mu^a \delta \zeta_a^\mu + \bar{\mu}_{\mu\nu}^a \delta I_a^{\mu\nu} \qquad (3.26)$$

in the Eulerian picture.

When the virtual displacement $\delta s$ is integrable, the virtual work will take the form:

$$\phi(\delta^1 \psi) = F_\mu \delta \xi^\mu + \bar{M}_{\mu\nu} \delta I^{\mu\nu} + \sigma_\mu^a \partial_a \delta \xi^\mu + \bar{\mu}_{\mu\nu}^a \partial_a \delta I^{\mu\nu} \; . \qquad (3.27)$$

When one applies the product rule for differentiation, this will become:

$$\phi(\delta^1 \psi) = (F_\mu - \partial_a \sigma_\mu^a) \delta \xi^\mu + (\bar{M}_{\mu\nu} - \partial_a \bar{\mu}_{\mu\nu}^a) \delta I^{\mu\nu} + \partial_a (\sigma_\mu^a \delta \xi^\mu + \bar{\mu}_{\mu\nu}^a \delta I^{\mu\nu}) \; , \qquad (3.28)$$

which we write in the form:

$$\phi(\delta^1 \psi) = j^* \phi(\delta \psi) + \partial_a \Pi^a(\delta \psi) \qquad (3.29)$$

with:

$$j^* \phi = (F_\mu - \partial_a \sigma_\mu^a) \, d\xi^\mu + (\bar{M}_{\mu\nu} - \partial_a \bar{\mu}_{\mu\nu}^a) \, dI^{\mu\nu} \; , \qquad (3.30)$$

$$\Pi^a = \sigma_\mu^a d\xi^\mu + \bar{\mu}_{\mu\nu}^a dI^{\mu\nu} \; . \qquad (3.31)$$

The *total virtual work* that is done on $s$ by the virtual displacement $\delta s$ is simply the integral of $\phi(\delta s)$ over $\mathcal{O}$ when it is pulled back by way of $s$:

$$\delta W[\delta s] = \int_\mathcal{O} \phi(\delta s) V_p \; , \qquad (3.32)$$

in which $V_p$ is the volume element on $\mathbb{R}^p$.

When the virtual displacement is integrable, this will become:

$$\delta W[\delta s] = \int_\mathcal{O} j^* \phi(\delta \psi) V_p + \int_{\partial \mathcal{O}} \Pi^a(\delta \psi) \# \partial_a \; , \qquad (3.33)$$

in which:

$$\# \partial_a = i_{\partial_a} V_p \qquad (3.34)$$

is then the volume element that is induced on the boundary of $\mathcal{O}$.

*e. Poincaré-invariant fundamental forms.* Previously, we found that in order to get the non-relativistic Cosserat equations of equilibrium, one had to focus on a particular type of fundamental 1-form, namely, ones that were invariant under the action of the group of Euclidian rigid motions. Now, we need to extend that condition to the Poincaré group, instead of the rigid motions.

First, one notes that $\phi$ is invariant under Poincaré transformations iff for every $\delta s$ that takes the form of a fundamental vector field for an infinitesimal Poincaré transformation, one has:

$$0 = L_{\delta s}\phi = i_{\delta s}d\phi + di_{\delta s}\phi . \tag{3.35}$$

Now, if $\phi$ is closed, as it is when it is $d\mathcal{L}$ for some Lagrangian density, then the first term will vanish. When will then remain is:

$$i_{\delta s}\phi = \phi(\delta s) = \text{constant} \tag{3.36}$$

for all such $\delta s$. Since $\phi(\delta s)$ will be the virtual work that is done by a relativistic "rigid" motion, with no loss of generality, one can make the constant equal to 0.

The condition that $\delta s$ must be the fundamental vector field for an infinitesimal Poincaré transformation amounts to the local condition that:

$$\delta \zeta^{\mu} = \text{const.}, \quad \delta I_{\nu}^{\mu} = \text{const.}, \qquad \delta \zeta_a^{\mu} = 0, \qquad \delta I_{\nu a}^{\mu} = 0. \tag{3.37}$$

This makes the virtual work take the form:

$$\phi(\delta s) = F_{\mu}\delta \zeta^{\mu} + \bar{M}_{\mu\nu}\delta I^{\mu\nu} , \tag{3.38}$$

which can be zero for all such $\delta s$ iff:

$$F_{\mu} = 0, \qquad \bar{M}_{\mu\nu} = 0. \tag{3.39}$$

When one goes back to the definition (3.22) of $\bar{M}_{\mu\nu}$, one sees that actually the object can still have a couple-moment acting on it, although it is of purely "internal" origin, namely:

$$M_{[\mu}^{\kappa}e_{\nu]\kappa} = -\sigma_{[\mu}^a x_{\nu]a} - \mu_{[\mu}^{\kappa a}e_{\nu]\kappa a} . \tag{3.40}$$

The fundamental 1-form then takes the form:

$$\phi = \sigma_{\mu}^a \zeta_a^{\mu} + (\mu_{\mu\nu}^a + \sigma_{[\mu}^a x_{\nu]})\mathcal{I}_a^{\mu\nu} , \tag{3.41}$$

and the virtual work that is done by an integrable virtual displacement is:

$$\phi(\delta^1 \psi) = -(\partial_a \sigma_{\mu}^a)\delta \zeta^{\mu} - (\partial_a \mu_{\mu\nu}^a + \sigma_{[\mu}^a x_{\nu]a})\delta I^{\mu\nu} + \partial_a(\sigma_{\mu}^a \delta \zeta^{\mu} + \mu_{\mu\nu}^a \delta I^{\mu\nu}) . \tag{3.42}$$

*f. The equations of motion/equilibrium.* In order to get the equations of motion/equilibrium, one must appeal to d'Alembert's principle of virtual work that the total virtual work must vanish for every integrable virtual displacement of $s$ that satisfies appropriate boundary-value conditions on $\partial\mathcal{O}$.

When one evaluates $\phi(\delta^1\psi)$ and integrates, one will get:

$$\delta W = \int_{\mathcal{O}} j^* \phi(\delta\psi) V_p + \int_{\partial\mathcal{O}} \Pi^a(\delta\psi) \# \partial_a \qquad (3.43)$$

for the total virtual work that is done by the virtual displacement.

As long as the boundary conditions on $\delta\psi$ make the boundary integral vanish in any case (e.g., variations vanishing on a fixed boundary or transversal variations on a free boundary), one can concisely express the resulting equations of motion/equilibrium as:

$$j^* \phi = 0. \qquad (3.44)$$

When one uses $\phi(\delta^1\psi)$ in the form (3.28), this yields the following system of first-order partial differential equations for the stress state in the object:

$$F_\mu = \partial_a \sigma_\mu^a, \qquad\qquad \bar{M}_{\mu\nu} = \partial_a \bar{\mu}_{\mu\nu}^a. \qquad (3.45)$$

When one goes back to the definitions for $\bar{M}_{\mu\nu}$ and $\bar{\mu}_{\mu\nu}^a$, one will see that the second set of equations takes the form:

$$M_{[\mu\nu]} = \partial_a \mu_{\mu\nu}^a. \qquad (3.46)$$

Although these are reasonable equations of motion/equilibrium, nonetheless, they are not the Cosserat equations. In order to get the latter equations, one must use $\phi(\delta^1\psi)$ for a Poincaré-invariant fundamental form, as in (3.42). The resulting equations of motion/equilibrium are then:

$$0 = \partial_a \sigma_\mu^a, \qquad\qquad 0 = \partial_a \bar{\mu}_{[\mu\nu]}^a + \sigma_{[\mu}^a x_{\nu]a}. \qquad (3.47)$$

As in Part I, if one wishes to put these equations into Cosserat form, as it is usually defined nowadays, one must make the further restriction that the object $\mathcal{O}$ is four-dimensional, so the matrix $x_a^\mu = \partial x^\mu / \partial \rho^a$ becomes invertible. When one multiplies both sets of equations by the inverse $\tilde{x}_\mu^a = \partial \rho^a / \partial x^\mu$ and uses the chain rule for differentiation, the resulting system of equations will take the form:

$$0 = \partial_\nu \sigma_\mu^\nu, \qquad\qquad 0 = \partial_\kappa \mu_{[\mu\nu]}^\kappa + \sigma_{[\mu\nu]}. \qquad (3.48)$$

In this form, the equations are defined purely on the deformed state and do not involve the parameterization of the object by $\mathcal{O}$.

**5. The free Dirac electron.** The Weyssenhoff fluid that we shall discuss in the next section is a relativistically-spinning fluid that represents an approximation to the tensorial form of the wave function of the Dirac electron. Thus, we shall first summarize the tensorial form of the Dirac electron in order to motivate the form of the Weyssenhoff fluid and then go on to address that fluid more specifically. As we shall see, the Dirac electron gives a physically-fundamental example of a relativistic Cosserat object.

By now, the standard (up to conventions) way of representing the wave function of the free electron/positron is by way of a differentiable function $\psi : \mathfrak{M}^4 \to \mathbb{C}^4$ and its Hermitian conjugate $\psi^\dagger$, or rather, its Dirac conjugate:

$$\overline{\psi} = \psi^\dagger \gamma^0.$$

Here, the matrices $\gamma_\mu$, $\mu = 0, 1, 2, 3$ represent the generators $\mathbf{e}_\mu$, $\mu = 0, 1, 2, 3$ of the Clifford algebra $\mathcal{C}(1, 3)$ over Minkowski space, and the matrices $\gamma^\mu$ represent the generators $\theta^\mu$ of the Clifford algebra over its dual space. Thus, the matrices are related by:

$$\gamma_\mu = \eta_{\mu\nu} \, \gamma^\nu = \pm \, \gamma^\mu.$$

Whereas $\gamma^0$ is Hermitian, the other three are anti-Hermitian. Furthermore, insofar as the gamma matrices define a faithful (i.e., injective) representation of the Clifford algebra $\mathcal{C}(1, 3)$ in the algebra of 4×4 complex matrices $M(4, \mathbb{C})$, they must satisfy:

$$\gamma_\mu \gamma_\nu + \gamma_\nu \gamma_\mu = 2 \eta_{\mu\nu}.$$

Since the Clifford algebra over any four-dimensional real vector space has real dimension $2^4 = 16$ and the matrix algebra of 4×4 complex matrices has real dimension 32, one sees that the representation of $\mathcal{C}(1, 3)$ by such matrices cannot be an isomorphism, except onto its image, which is then a proper sub-algebra of that matrix algebra.

Although the representation of $\mathcal{C}(1, 3)$ is not unique, a common choice of matrices is the Dirac representation [**4, 5**]:

$$\gamma^0 = \begin{bmatrix} I & 0 \\ 0 & -I \end{bmatrix}, \qquad \gamma^i = \begin{bmatrix} 0 & \sigma^i \\ -\sigma^i & 0 \end{bmatrix},$$

in which the matrices $\sigma^i$, $i = 1, 2, 3$ are the Pauli matrices that define a basis for the Lie algebra $\mathfrak{su}(2)$.

The equations that must be satisfied by $\psi$ and $\overline{\psi}$ are then the two forms of the *Dirac equation*:

$$i\partial\!\!\!/\,\psi = \kappa\psi, \qquad\qquad i\partial\!\!\!/\,\overline{\psi} = -\,\kappa\overline{\psi}. \qquad\qquad (4.1)$$

In this, the constant $\kappa = mc/\hbar$ represents the Compton wavelength when $m$ is the rest mass of the electron/positron. We have also defined the Dirac operator by:

$$\vec{\partial\!\!\!/} = \gamma^\mu \frac{\partial}{\partial x^\mu}.$$

The way that one couples this wave function to an external electromagnetic field, which one represents by its potential 1-form $A = A_\mu \, dx^\mu$ for some choice of gauge is by way of "minimal electromagnetic coupling," which amounts to the replacement of $\vec{\partial\!\!\!/}$ by:

$$\vec{\nabla\!\!\!\!/} = \vec{\partial\!\!\!/} + eA,$$

in which $e$ is then the charge of the electron.

The Dirac equation – in either its free or electromagnetically-coupled form – can be obtained from an action functional, whose Lagrangian density takes the form:

$$\mathcal{L} = \frac{i\hbar c}{2}(\overline{\psi}\,\vec{\partial\!\!\!/}\,\psi - \overline{\psi}\,\overleftarrow{\partial\!\!\!/}\,\psi + 2i\kappa\overline{\psi}\psi) \tag{4.2}$$

in the former case. In this expression, the arrows over the Dirac operator indicate which wave function they act on. One notes that whenever the wave function $\psi$ and its Dirac conjugate satisfy the Dirac equations, the Lagrangian density will vanish.

When one applies Noether's theorem to this Lagrangian density, one will see that each infinitesimal symmetry of the action functional, which include the symmetries of the Lagrangian density, will correspond to a divergenceless vector field on $\mathfrak{M}^4$ that one can regard as the conserved current that is associated with that symmetry. The vanishing of those divergences will then give the equations of motion for the physical observables that they describe.

The most immediate symmetry of $\mathcal{L}$ is under phase transformations, which replace $\psi$ with $e^{-i\phi}\psi$ and $\overline{\psi}$ with $e^{i\phi}\overline{\psi}$. The variations of the fields $\psi$ and $\overline{\psi}$ then take the form:

$$\delta\psi = -i\phi\,\psi, \quad \delta\overline{\psi} = i\phi\,\overline{\psi}.$$

The Noether current that corresponds to this takes the form:

$$j^\mu = \frac{\partial \mathcal{L}}{\partial \psi_{,\mu}}\delta\psi + \frac{\partial \mathcal{L}}{\partial \overline{\psi}_{,\mu}}\delta\overline{\psi} = \hbar c\,\overline{\psi}\,\gamma^\mu\psi \equiv \hbar c\,S^\mu. \tag{4.3}$$

The vector field $\mathbf{S} = S^\mu\,\partial_\mu$ that we have defined then has the property that its Minkowski norm $\rho$:

$$\rho^2 = \eta_{\mu\nu}\,S^\mu\,S^\nu = \frac{1}{\hbar^2 c^2}\,j^2 \tag{4.4}$$

is real and can thus represent a matter density.

One can also define a time-like unit-speed four-velocity vector field $\mathbf{u}$ using $\mathbf{j} = j^\mu\, \partial_\mu$ by way of:

$$\frac{1}{\hbar}\mathbf{j} = \rho\mathbf{u} \ . \tag{4.5}$$

The conservation law that is associated with the current $\mathbf{j}$, namely, the vanishing of div $\mathbf{j}$, then gives the conservation law:

$$\partial_\mu(\rho u^\mu) = 0, \tag{4.6}$$

which then represents the conservation of whatever physical quantity – e.g., number, mass, or charge – is associated with the density $\rho$.

The symmetry of $\mathcal{L}$ under an infinitesimal translation $v^\mu$ involves a variation of the field $\psi$ that takes the form:

$$\delta\psi^a = \mathrm{L}_v\psi^a = \frac{\partial\psi^a}{\partial x^\mu}v^\mu \ .$$

The associated current:

$$P^\mu(v) \ = T_\nu^{\ \mu}\, v^\nu$$

is the linear momentum density of the wave, in which:

$$T_\nu^{\ \mu} = \mathcal{L}\,\delta_\nu^\mu - \pi_a^\mu\,\psi_{,\nu}^a - \overline{\pi}_a^\mu\,\overline{\psi}_{,\nu}^a \tag{4.7}$$

is the canonical energy-momentum tensor, and in this expression we have introduced the conjugate momentum:

$$\pi_a^\mu = \frac{\partial\mathcal{L}}{\partial\psi_{,\mu}^a}$$

and its analogous complex conjugate $\overline{\pi}_a^\mu$. Note that, in general, the canonical energy-momentum tensor does not have to be symmetric when one lowers the upper index, and in fact:

$$T_{[\mu,\,\nu]} = -\,\pi_{[\mu|a}\,\psi_{\nu]}^a - \overline{\pi}_{[\mu|a}\,\overline{\psi}_{\nu]}^a \ . \tag{4.8}$$

The conservation of energy-momentum then takes the form:

$$0 = \partial_\mu T_\nu^{\ \mu} \ . \tag{4.9}$$

When one uses the Dirac Lagrangian, the energy-momentum tensor will take the specific form:

$$T_\nu^{\ \mu} = \frac{\hbar c}{2}(\overline{\psi}\,\gamma^\mu\partial_\nu\psi - \partial_\nu\overline{\psi}\,\gamma^\mu\psi) \ , \tag{4.10}$$

which will then make:

$$T_{[\mu,\,\nu]} = \frac{\hbar c}{2}(\overline{\psi}\,\gamma_{[\mu}\partial_{\nu]}\psi - \partial_{[\nu}\overline{\psi}\,\gamma_{\mu]}\psi)\,. \tag{4.11}$$

We now discuss the conserved currents that are associated with infinitesimal symmetries under the action of Poincaré group on $\mathcal{L}$. The variation of the field $\psi$ that one defines for every infinitesimal Lorentz symmetry $\omega_\nu^\mu$ takes the form:

$$\delta\psi^a = (\mathfrak{D}_{b\mu}^{a\nu}\omega_\nu^\mu)\psi^b\,,$$

in which the symbol $\mathfrak{D}_{b\mu}^{a\nu}$ is really the matrix of the linear algebra homomorphism $\mathfrak{D}$: $\mathfrak{so}(1,3) \to \mathfrak{gl}(4,\mathbb{C})$ that represents the infinitesimal Lorentz transformation $\omega_\nu^\mu$ by way of the 4×4 complex matrix:

$$\mathfrak{D}_b^a(\omega) = \mathfrak{D}_{b\mu}^{a\nu}\omega_\nu^\mu\,.$$

In the present case, one has:

$$\mathfrak{D}_\mu^\nu = \tfrac{1}{8}\,[\gamma_\mu\,,\,\gamma^\nu]\,.$$

in which we have suppressed the *a-b* indices, which belong to the gamma matrices.

The Noether current that is associated with the infinitesimal Lorentz symmetry $\omega_\nu^\mu$ then takes the form:

$$M^\mu(\omega) = M_\nu^{\lambda\mu}\omega_\lambda^\nu\,,$$

in which:

$$M_\nu^{\lambda\mu} = L_\nu^{\lambda\mu} + S_\nu^{\lambda\mu}\,,$$

is the *total angular momentum tensor*, with:

$$L_\nu^{\lambda\mu} = -\,T_{[\nu}^\mu x^{\lambda]}\,, \qquad S_\nu^{\lambda\mu} = \mathfrak{D}_{b\nu}^{a\mu}\frac{\partial\mathcal{L}}{\partial\psi_{,\lambda}^a}\psi^b\,. \tag{4.12}$$

The former tensor $L_\nu^{\lambda\mu}$ represents the *orbital* angular momentum that is due to the energy-momentum tensor $T_\nu^\mu$ acting at a distance from the origin, whose displacement vector is then $\mathbf{r} = x^\mu\,\partial_\mu$. The latter one $S_\nu^{\lambda\mu}$ represents the *intrinsic* angular momentum – or *spin* – and one sees that its origin has as much to with the type of representation that one uses for the $\mathfrak{iso}(3,1)$ as it does with actual dynamical properties of $\psi$, which enter in by way of the wave function $\psi$ and its conjugate momentum.

The conservation of total angular momentum takes the general form:

$$0 = \partial_\mu M_\nu^{\lambda\mu} = \partial_\mu S_\nu^{\lambda\mu} - T_{\quad\nu]}^{[\lambda}\,. \tag{4.13}$$

When one specifies the Dirac Lagrangian and the representation of $\mathfrak{so}(1, 3)$ by gamma matrices, the spin tensor will initially take the form:

$$S_\nu^{\lambda\mu} = -\frac{\hbar c}{8}\overline{\psi}(\gamma^\mu\gamma^\lambda\gamma_\nu - \gamma_\nu\gamma^\lambda\gamma^\mu)\psi \,. \tag{4.14}$$

However, since:

$$S_\nu^{\mu\mu} = 0,$$

one can also express it the form:

$$S_\nu^{\lambda\mu} = -\frac{\hbar c}{4}\overline{\psi}(\gamma^\mu\gamma^\lambda\gamma_\nu)\psi \,. \tag{4.15}$$

We summarize the conservation laws that we have obtained up to now:

$$0 = \partial_\mu(\rho u^\mu), \quad 0 = \partial_\mu T_\nu^\mu, \qquad 0 = \partial_\mu S_\nu^{\lambda\mu} - T^{[\lambda}_{\cdot\nu]} \,. \tag{4.16}$$

The last one can also be stated in the form:

$$T_{[\lambda\nu]} = \partial_\mu S_{\lambda\nu}^\mu \,, \tag{4.17}$$

which then says that the asymmetry in the energy-momentum tensor is due to the divergence of the intrinsic angular momentum tensor.

Although the conservation laws (4.16) clearly have the form of the relativistic Cosserat equations (3.48), one should point out that they are actually generic to the conservation laws that one obtains from a field Lagrangian whose action functional is invariant under the Poincaré group. The contribution of the Cosserats was to consider the case in which $\partial_\mu S_\nu^{\lambda\mu}$ was non-vanishing, so the tensor $T_\nu^\mu$ became asymmetric when an index was lowered or raised.

One can also deduce the following constraints for the covelocity 1-form and the intrinsic angular momentum tensor:

$$u^2 = c^2, \qquad u_\mu S_\nu^\mu = 0, \tag{4.18}$$

in which we have defined the intrinsic angular momentum tensor by:

$$S_\nu^\mu = u_\lambda S_\nu^{\lambda\mu}, \tag{4.19}$$

which will then make:

$$S_\nu^{\lambda\mu} = u^\lambda S_\nu^\mu \,. \tag{4.20}$$

Although we shall not go into all of the details at present (see, however, Halbwachs [6]), we point out that the canonical energy-momentum tensor can be put into the form:

$$T_\nu^\mu = g_\nu u^\mu + \frac{\hbar c}{2} \partial_\nu A \ \hat{S}^\mu + S_\lambda^\mu \partial_\nu u^\lambda . \qquad (4.21)$$

In this expression, one sees the introduction of an energy-momentum 1-form $g = g_\nu \, dx^\nu$ that is distinguished by the fact that it does not have to collinear with the covelocity 1-form $u = u_\nu \, dx^\nu$, but can include a transverse momentum contribution that originates in the non-vanishing spin of the wave function.

The angle $A$ that was introduced by Takabayasi [7] comes from first expressing $\rho^2$ in the form:

$$\rho^2 = \Omega^2 + \hat{\Omega}^2 , \qquad (4.22)$$

and then expressing the two parameters $\Omega$ and $\hat{\Omega}$ thus-defined in the form:

$$\Omega = \rho \cos A, \quad \hat{\Omega} = \rho \sin A. \qquad (4.23)$$

The vector field $\hat{S}^\mu$ makes the bivector field $\mathbf{u} \wedge \hat{\mathbf{S}}$ Poincaré dual to the 2-form $S = \frac{1}{2} S_{\mu\nu} \, dx^\mu \wedge dx^\nu$:

$$S = \#(\mathbf{u} \wedge \hat{\mathbf{S}}) \qquad (S_{\mu\nu} = \frac{1}{2} \varepsilon_{\mu\nu\kappa\lambda} u^\kappa \hat{S}^\lambda ). \qquad (4.24)$$

According to Takabayasi [7], the last two terms on the right-hand side of equation (4.21) give one a heat current:

$$- c^2 \, q^\mu = \left( \frac{\hbar c}{2} \partial_\nu A \ \hat{S}^\mu + S_\lambda^\mu \partial_\nu u^\lambda \right) u^\nu \qquad (4.25)$$

and an internal stress tensor:

$$\theta_\nu^\mu = \frac{\hbar c}{2} \partial_\nu A \ \hat{S}^\mu + S_\lambda^\mu \partial_\nu u^\lambda + \frac{\hbar}{2} \dot{A} \hat{S}^\mu u_\nu + \frac{1}{c^2} S_\lambda^\mu \dot{u}^\lambda u_\nu . \qquad (4.26)$$

The latter tensor suggests that there is an internal pressure of:

$$\tfrac{1}{3} \theta_\mu^\mu = \frac{\hbar}{2} \partial_\mu A \ \hat{S}^\mu + \tfrac{1}{2} S^{\mu\nu} (\partial_\mu u_\nu - \partial_\nu u_\mu) \qquad (4.27)$$

that acts in the wave.

One can also associate a proper mass density:

$$\mu_0 = m_0 \rho \cos A + \frac{1}{3c^2} \theta_\mu^\mu . \qquad (4.28)$$

Thus, in addition to the usual contribution $m_0 \rho$ from rest mass, one finds a contribution from the internal pressure.

**6.  The Weyssenhoff fluid.**  The abstraction that Weyssenhoff [**8**] made (the contribution of his former student Raabe seems to have been posthumous to the seminal articles) was to start with a relativistic spinning fluid in a region $R$ of Minkowski space. The fluid is defined by a rest mass density $\rho_0$, a flow velocity four-vector field $\mathbf{u} = u^\mu \, \partial_\mu$, an energy-momentum density covector field $g = g_\mu \, dx^\mu$, and a spin density tensor field $s_\mu^{\;\nu}$, which one represents as a 0-form with values in $\mathfrak{so}(1, 3)^*$.  The support of $\rho_0$ is $R$, by the definition of $R$, and the supports of the other fields are then contained in $R$.  As we shall see, the density $\rho_0$ can actually be obtained from $g$, so it cannot be specified independently of $g$.

Since the basic equations of motion/equilibrium are still the conservation laws for mass, energy-momentum and angular momentum that one uses for relativistic Cosserat media, the Weyssenhoff fluid represents an example of a relativistic Cosserat medium that is obtained by specializing the form of the stress-energy-momentum and couple-stress-angular-momentum tensor to a particular form that was motivated by Frenkel's relativistic equations of motion for a spinning point-like electron, as well as the tensorial form of the Dirac electron.

*a.  The basic fields.*  The basic fields are then subjected to various constraints that basically grow out of the properties of the Dirac electron.  First of all, the flow velocity vector field is assumed to be time-like and proper-time parameterized, so one must have:

$$u^2 = c^2, \qquad u^\mu = \frac{dx^\mu}{d\tau}. \tag{5.1}$$

However, no immediate constraint is imposed upon the *kinematical vorticity* 2-form $\Omega_k$ of $\mathbf{u}$, which is defined to be the exterior derivative of the covelocity 1-form $u = u_\mu \, dx^\mu$:

$$\Omega_k \equiv du = -\tfrac{1}{2}(\partial_\mu u_\nu - \partial_\nu u_\mu) \, dx^\mu \wedge dx^\nu. \tag{5.2}$$

One can decompose this 2-form into time-space form as:

$$\Omega_k = -\tfrac{1}{2}(\partial_0 u_i - \partial_i u_0) \, dx^0 \wedge dx^i - \tfrac{1}{2}(\partial_i u_j - \partial_j u_i) \, dx^i \wedge dx^j. \tag{5.3}$$

One can then define the *kinematical compressibility* of the fluid by the divergence of the flow velocity four-vector field:

$$\chi_k \equiv \partial_\mu u^\mu. \tag{5.4}$$

As for the energy-momentum 1-form $g$, it is not assumed to be necessarily collinear with the covelocity 1-form, but can include a transverse momentum contribution $\pi$, so one can represent it in the form:

$$g = \rho_0 \, u + \pi. \tag{5.5}$$

in which $\rho$ is the mass density in the natural frame.  By assumption, since $\pi$ is transverse to $u$, one must have:

$$\eta(u, \pi) = 0. \tag{5.6}$$

One can define two rest mass densities corresponding to $g$, namely, the *rest mass density of inertia* $\rho_0$ and the *rest mass density of momentum* $\mu_0$, which are defined by:

$$\rho_0 \, c^2 \equiv g(\mathbf{u}) = g_\mu \, u^\mu, \qquad \mu_0^2 \, c^2 = g(\mathbf{g}) = g_\mu \, g^\mu = \rho_0^2 \, c^2 + \pi^2. \tag{5.7}$$

No immediate constraint is imposed upon the *dynamical vorticity* 2-form $\Omega_d$, which is the exterior derivative of the energy-momentum 1-form:

$$\Omega_d \equiv dg = -\tfrac{1}{2} \, (\partial_\mu \, g_\nu - \partial_\nu \, g_\mu) \, dx^\mu \wedge dx^\nu. \tag{5.8}$$

The time-space form of this is:

$$\Omega_d = -\tfrac{1}{2} \, (\partial_0 \, g_i - \partial_i \, g_0) \, dx^0 \wedge dx^i - \tfrac{1}{2} \, (\partial_i \, g_j - \partial_j \, g_i) \, dx^i \wedge dx^j. \tag{5.9}$$

The *dynamical compressibility* of the fluid is the divergence of the energy-momentum vector field that is metric dual to the 1-form $g$:

$$\chi_d = \partial_\mu \, g^\mu = \mathbf{u}\rho_0 + \rho_0 \chi_k + \partial_\mu \, \pi^\mu. \tag{5.10}$$

It is tempting to associate the vanishing of $\chi_d$ with the conservation of rest mass density. However, since we now have two different types of rest mass density to deal with, namely, $\rho_0$ and $\mu_0$, we must consider the conservation of each separately, which we will do after we have discussed the equations of motion more thoroughly.

Finally, the spin density $s_\mu^\nu$ is subjected to the *Frenkel constraint*, that it be "purely magnetic":

$$s_\mu^\nu u^\mu = 0. \tag{5.11}$$

The reason that one thinks of this as a magnetic constraint is based in the idea that when Frenkel [9] was first devising his relativistic theory of the electrodynamics of spinning charges, he was imagining that that the spin would be coupled to an magnetic moment by means of a scalar factor, such as the Bohr magneton for the electron, while the electric dipole moment would vanish in the rest system, which would be consistent with the assumption of a spherical charge distribution, as well as the experiments.

Up to this point, we have been duplicating the properties of the Dirac electron that we deduced in the previous section. However, the dynamical equations that we defined above must be specialized to a particular choice of stress-energy-momentum tensor $T_\mu^\nu$ and intrinsic angular-momentum tensor $S_\mu^{\nu\lambda}$, while the orbital angular momentum follows from the form of $T_\mu^\nu$. The choice that Weyssenhoff made was essentially a simplification of the case for the Dirac electron that one obtains by eliminating the internal stress contribution. What one will then have left is:

$$T = g \otimes \mathbf{u}, \qquad S = s \otimes \mathbf{u}; \tag{5.12}$$

i.e.:

$$T_\mu^\nu = g_\mu u^\nu, \qquad\qquad S_\mu^{\nu\lambda} = s_\mu^\nu u^\lambda.$$

One can also put $T$ into the longitudinal-transverse form:

$$T = \rho_0 u \otimes \mathbf{u} + \pi \otimes \mathbf{u} \qquad (T_\mu^\nu = \rho_0 u_\mu u^\nu + \pi_\mu u^\nu). \tag{5.13}$$

$T$ is not generally symmetric, and since it does not depend upon the rate of deformation of the fluid, one assumes that the fluid that one is defining is inviscid. Indeed, calling this form of matter a "fluid" to begin with is somewhat less than precise.

In the co-moving system for $u$, one has $\mathbf{u} = (c, 0, 0, 0)^T$ and $u = (c, 0, 0, 0)$, which makes the first part of the tensor take the form of something whose only non-zero component is $\rho_0 u_0 u^0 = \rho_0 c^2$; i.e., the rest mass-energy of a non-spinning distribution. If $\pi_\mu = (0, \pi_1, \pi_2, \pi_3)$ then the second part will be non-vanishing only for $\pi_j u^0 = c\pi_j$, with $j = 1, 2, 3$. Thus, the form of the stress-energy-momentum tensor in the co-moving frame will be:

$$T_\mu^\nu = \left[ \begin{array}{c|c} \rho_0 c^2 & c\pi_j \\ \hline 0 & 0 \end{array} \right]. \tag{5.14}$$

The first part of $T$ is the purely kinetic contribution to the stress, energy, and momentum, and becomes symmetric when one lowers the upper index, while the second part accounts for the non-vanishing anti-symmetric part of that tensor field, since:

$$T_{(\mu\nu)} = \rho_0 u_\mu u_\nu + \tfrac{1}{2} (\pi_\mu u_\nu + \pi_\nu u_\mu), \quad T_{[\mu\nu]} = \tfrac{1}{2} (\pi_\mu u_\nu - \pi_\nu u_\mu). \tag{5.15}$$

Thus, the term in the symmetric part, by itself, describes a cloud of dust that moves with a relativistically-meaningful velocity relative to the observer, while the second term represents the contribution from the transverse momentum. The anti-symmetric part of $T$, when the upper index is lowered, then defines a 2-form that takes the form:

$$\pi \wedge u = c\pi_i \, dx^i \wedge dx^0 + \frac{c}{2} (\pi_i u_j - \pi_j u_i) \, dx^i \wedge dx^j. \tag{5.16}$$

In the comoving frame, it then becomes purely "electric" in character.

The trace of the tensor $T_\mu^\nu$ will then become:

$$T_\mu^\mu = g_\mu u^\mu = \rho_0 c^2. \tag{5.17}$$

*b. The equations of motion.* With the above definitions, the law of conservation of energy-momentum now takes the form:

$$0 = d_\tau \, g_\mu = \frac{dg_\mu}{d\tau} + \chi_k \, g_\mu, \qquad\qquad (5.18)$$

in which, following Weyssenhoff, we have defined the *density derivative* operator by:

$$d_\tau f = \partial_\nu (f \, u^\nu) = \frac{df}{d\tau} + \chi_k f. \qquad\qquad (5.19)$$

Its significance is derived from the fact that when one integrates any density $f(t, x^i)$ over a space-like cross-section $\Sigma(\tau)$ of the world-tube that is swept out by the support of the fluid in its motion, one will obtain a function $F(\tau)$ with the property that:

$$\frac{dF}{d\tau} = \int_{\Sigma(\tau)} (d_\tau f) \, \#\mathbf{u} \qquad\qquad (\#\mathbf{u} = i_\mathbf{u} V).$$

The density derivative does not behave like a derivation precisely with regard to the product rule, but gives:

$$d_\tau (fg) = (d_\tau f)g + f \, \frac{dg}{d\tau}. \qquad\qquad (5.20)$$

When one takes the divergence of the orbital angular momentum, one will obtain:

$$\partial_\lambda L^{\mu\lambda}_{\;\;\nu} = t^{[\mu}_{\;\cdot \nu]} - t^{[\cdot\nu}_{\;\mu]} = \pi_\mu \, u^\nu - \pi_\nu \, u^\mu. \qquad\qquad (5.21)$$

The divergence of the intrinsic angular momentum tensor now takes the form:

$$d_\tau s^\mu_\nu = \partial_\lambda (s^\mu_\nu u^\lambda) = (\partial_\lambda s^\mu_\nu) u^\lambda + \chi_k s^\mu_\nu. \qquad\qquad (5.22)$$

Putting the orbital and spin contributions together, one can give the conservation of angular-momentum the Weyssenhoff form:

$$d_\tau s^\mu_\nu + \pi^\mu \, u_\nu - \pi_\nu \, u^\mu = 0, \qquad\qquad (5.23)$$

which again has the Cosserat form.

One can now see the coupling of the transverse momentum to the time derivative of spin by contracting the last equation with $u^\nu$:

$$\pi^\mu = \frac{1}{c^2} (d_\tau s^\mu_\nu) u^\nu = -\frac{1}{c^2} s^\mu_\nu \frac{du^\nu}{dx}. \qquad\qquad (5.24)$$

In the last step, we have differentiated the Frenkel constraint in order to replace $(d_\tau s^\mu_\nu) u^\nu$ with $-s^\mu_\nu du^\nu / d\tau$:

$$0 = d_\tau(s_\nu^\mu u^\nu) = (d_\tau s_\nu^\mu)u^\nu + s_\nu^\mu \frac{du^\nu}{d\tau}.$$

If one substitutes this into the expression for $g_\nu$ then one will get the new form for the energy-momentum 1-form:

$$g_\nu = \rho_0 \, u_\nu \, -\frac{1}{c^2} s_{\mu\nu} \, a^\nu \, , \qquad (5.25)$$

in which $a^\nu = du^\nu / d\tau$ is the rest acceleration.

**7.  Discussion.** – The work of the Cosserat brothers ([1]) on the description of the equilibrium and motion of deformable bodies that admitted internal couple-stresses by means of action functionals that were invariant under Euclidian rigid motions was originally known to only a few enlightened figures in theoretical mechanics and mathematics.  It was resurrected in the 1950's by a school of researchers that largely worked in non-relativistic continuum mechanics, although Hehl [**10**] did suggest at that time that the concept of a Cosserat medium might be analogous to the geometry of space-time when one extended general relativity from Riemannian geometry to Riemann-Cartan geometry by the addition of non-vanishing torsion.  However, the details of how one would generalize a Cosserat medium from non-relativistic to relativistic continuum mechanics was never addressed in a thorough manner, although it was implicitly included in Pommaret's treatment [**11**] of Cosserat media with the methods of Lie groupoids and Lie equations.  Some of the reason for that might be that most of the literature on the Cosserat equations (except for the Pommaret book and a paper by Schaefer that is also cited in Part I) started with those equations as given *a priori* without taking the time to derive those equations from first principles, while the derivation that was given by the Cosserat's was hampered by the fact that Noether's theorem was conspicuous by its absence, due to the fact the Cosserat book was published several years before Noether's paper on invariant variational problems.  Interestingly, when Sudria (see Part I) revisited the theory of Euclidian-invariant action functionals more than a decade after Noether's theorem was published, he made no mention of it, either.  Hence, one should never assume that merely because some fundamental result was in print during some era of history, the entire research community was immediately aware of its existence.  Indeed, in the thirty-five years that followed the publication of Mendel's laws of genetics, his paper was cited only three times, and apparently Charles Darwin was unaware of its existence at the time that he wrote his landmark treatise on the evolution of species.

In the eyes of physics, it is useful to know that nature has given us a concrete example of a relativistic Cosserat medium in the form of the free Dirac electron.  In the absence of empirically-defined examples to give it tangible roots, theoretical physics has a natural tendency to simply follow the path of mathematical generalization without stopping to apply the successive generalizations to physical phenomena.  Tangible examples that one can examine in a terrestrial laboratory can serve to keep the scope of the mathematics focused upon the specializations that are most useful to the modeling of those examples.

---

([1])  For the classical i.e., non-relativistic) references on Cosserat media, see the bibliography to Part I [**1**].

## References ([*])

---

([*])  References marked with an asterisk are available in English translation at the author's website: neo-classical-physics.info.